\documentclass[aps,pra,twocolumn,groupedaddress]{revtex4}
\usepackage[dvips]{graphicx}
\usepackage{float}
\usepackage{amsmath}
\usepackage{color}

\begin{document}


\title{Breakdown of macroscopic quantum self-trapping in coupled mesoscopic one dimensional Bose gases}


\author{Rafael Hipolito}
\email[]{rafa999@buphy.bu.edu}
\affiliation{Department of Physics, Boston University, Boston, MA 02215}
\author{Anatoli Polkovnikov}
\affiliation{Department of Physics, Boston University, Boston, MA 02215}


\date{\today}

\begin{abstract}
Two coupled BECs with a large population imbalance exhibit macroscopic quantum self-trapping (MQST) if the ratio of interaction energy to tunneling energy is above a critical value.  Here we investigate effect of quantum fluctuations on MQST. In particular, we analyze the dynamics of a system of two elongated Bose gases prepared with a large population imbalance, where due to the quasi one dimensional character of the gas we no longer have true long range order, and the effect of quantum fluctuations is much more important. We show that MQST is possible in this system, but even when it is achieved it is not always dynamically stable. Using this instability one can construct states with sharply peaked momentum distributions around characteristic momenta dependent on system parameters.  Our other finding is the nonmonotonic oscillating dependence of the decay rate of the MQST on the length of the wires. We also address the interesting question of thermalization in this system and show that it occurs only in sufficiently long wires.
\end{abstract}

\pacs{}

\maketitle

\section{Introduction}
The Josephson effect~\cite{giovanazzi2000,levy2007,zapata1998} and macroscopic quantum self
trapping (MQST)~\cite{raghavan1999,albiez2005} have been studied extensively both theoretically
and experimentally in the context of two coupled BECs. MQST has been observed experimentally
in a single bosonic Josephson junction~\cite{albiez2005}.  In these cases, one considers a two
well system prepared with a large initial population imbalance and investigates the dynamics.
The occurrence of MQST can be understood by mapping the system into an equivalent classical
nonrigid pendulum system, with the pendulum length depending on angular momentum~\cite{raghavan1999}. Angular momentum in the pendulum maps to population
difference in the coupled BEC, and likewise angular displacement maps to the relative phase.
Then the Josephson effect is analogous to small oscillations of the pendulum around equilibrium
(where the system has low energy), and MQST corresponds to the pendulum making full revolutions
(high energy). In the high energy pendulum case, the average angular momentum is nonzero and the
angle monotonically increases in time.  This corresponds to nonzero average population difference
and a running phase in the BEC case, which defines MQST.

MQST can be understood using the following qualitative considerations. Consider the case where initially all atoms occupy one well. The transition to MQST is a nonlinear effect governed by the parameter $\lambda=J/\mu$, where $\mu$ is the chemical potential in the full well (which is approximately equal to its interaction energy per atom), and $J$ is the tunneling energy between the two wells.  Self trapping occurs when $\lambda$  is smaller than some critical value of the order of unity. The transition to MQST is understood in terms of a simple energy argument. When the parameter $\lambda$ is small, corresponding to large interaction energy, the initial state with a large population imbalance has a very high energy. We might expect the system to dynamically equilibrate and the imbalance to disappear.  However, there is a kinematic barrier for this to occur since the interaction energy has to be transferred to the kinetic energy of atoms. If the condensate depletion is small and the dynamics is effectively constrained to the condensate modes (this is the case e.g. for weakly interacting 3D condensates) then the only source of the kinetic energy is the tunneling term. For small $\lambda$ the gain in hopping energy cannot make up for the loss in interaction energy thus there is a phase space restriction to dynamics. The best the system can do is a partial tunneling of atoms, and the system is constrained to remain imbalanced.  This is analogous to giving the pendulum a strong enough initial kick to make full revolutions, where the angular momentum remains nonzero at all times. Here we have a large \emph{average} angular momentum, where deviations from this average (due to the presence of the external gravitational field) oscillate in time with a high frequency and small amplitude.  It is these angular momentum deviations in the pendulum case that are analogous to the high frequency small amplitude tunneling of atoms in the bosonic Josephson junction in MQST~\cite{raghavan1999,albiez2005}. Finally, let us briefly mention that we can also understand MQST through an analogy with the emergence of second order atom tunneling in double well systems due to superexchange (see Refs.~\cite{trotzky_07, trotzky_08} and the endnote~\endnote{If two atoms occupying a double well are driven away from equilibrium then at small interactions one observes single frequency Josephson oscillations while at strong interactions one observes oscillations described by two frequencies. The first high frequency small amplitude oscillation describes single atom virtual tunneling to the unoccupied well, and the second, low frequency oscillation, describes simultaneous co-tunneling of two atoms. As one increases the number of atoms per well the second collective tunneling mode becomes exponentially suppressed and the first mode describes MQST.})
 
The explanation above of MQST heavily relies on the assumption that the condensate remains coherent at all times. This assumption is essential since the interaction energy can be also transferred to the kinetic energy of the non-condensate modes within each well. As we mentioned this process is kinematically suppressed if we are dealing with three-dimensional condensates where the depletion is small and quantum fluctuations required to excite such modes are negligible. In one-dimensional systems, on the contrary, the situation is quite different.
Quantum fluctuations are very important. In equilibrium they result in destruction of the condensate in the thermodynamic limit even at zero temperature~\cite{giamarchi_book}. One can expect that these fluctuations will also affect dynamics. This is indeed the case in a different setup, where it was shown that quantum fluctuations were responsible for destruction of the effective spin echo in one-dimensional condensates~\cite{widera_07}. As we will explain in this paper the noncondensate modes crucially affect the dynamics expected from the simplified self-trapping picture.

In this paper we will consider the two wire system, where all atoms are initially loaded into one wire and then the wires are suddenly coupled. Such a system was recently realized experimentally~\cite{bloch_private} and analyzed theoretically in a different regime~\cite{huber_09}. We assume that initially the atoms are in the ground state of a single full wire, and then the system is quenched by suddenly coupling the wires by the tunneling term.  In this work we will focus on the self-trapping regime where the tunneling is much smaller than the interaction energy per atom in the full wire. This initial state then describes a highly nonequilibrium situation and we can expect interesting dynamics. We then ask, is it possible to have MQST, and is it stable?  If not, what is the mechanism of MQST decay?  Does the system eventually equilibrate, and if so, at what time scales?

To tackle the problem we will use the idea that in the self-trapping regime we have clear separation of time scales in the dynamical process. The high frequency small amplitude oscillations expected from zero dimensional analogy with Josephson junction~\cite{raghavan1999,albiez2005}, will then serve as a source term for slow dynamical processes occurring at much longer time scales. The whole dynamics is then analogous to that in an externally driven system by a periodic modulation with frequency much larger than the natural transition energy in the system. Then one can effectively integrate out high frequency modes by averaging dynamics over short time scales and get coarse grained effective description of the slow degrees of freedom (see e.g. Ref.~\cite{landaulifshitzmec} for illustrations of this idea in the context of classical mechanics).  This analogy allows one to construct an effective ``low energy'' description of the dynamics in our system describing slow degrees of freedom. The analogy with externally driven systems is not perfect, however, since in our situation there is a feedback from the slow modes affecting the fast coherent dynamics. As we will show this process inhibits original fast oscillations and leads to decoherence and eventual thermalization in the system. We have to rely on numerical simulations to describe this process. Nevertheless at intermediate times the picture where one treats high frequency degrees of freedom separately providing external source to low frequency fields gives very accurate quantitative description of the dynamics and allows us to derive analytical expressions describing key instabilities in the system destroying self-trapping.

Another important result of the present analysis is the evidence of long-time thermalization in the system.
We note that the dilute interacting 1D Bose gas is well described by a delta function interaction, which is integrable in both classical and quantum descriptions~\cite{lieb1963}, and therefore does not thermalize in the usual sense~\cite{rigol2007}.  However, the coupling between the two 1D wires breaks integrability and
we might expect the system to thermalize. The issue of thermalization in weakly nonintegrable systems is far from clear (see e.g. Ref.~\cite{rigol2009}).  In a classical system, the KAM theorem states that adding a weak perturbation to an integrable system precludes the system from thermalizing (see Refs.~\cite{tabor1989,arnold1978}; for the original KAM papers see Refs.~\cite{arnold1963,kolmogorov1957,moser1962}). It is very little known what happens in quantum weakly integrable systems. In this work we address this issue numerically for our particular problem and give evidence that like in the classical systems thermalization may or may not occur in the analyzed setup depending on the choice of parameters. We note that the similar conclusions of thermalization due to breaking integrability were reached in Ref.~\cite{mazets} about thermalization in a single wire due to virtual excitations of higher radial modes. This paper concluded that thermalization should occur for arbitrarily weak coupling to virtual excitations. However, experiments show that non-thermalizable dynamics in quasi one-dimensional systems of bosons can be very robust~\cite{kinoshita}, so the whole picture of thermalization in nearly integrable quantum systems remains quite controversial.

Microscopically the system we are considering is described by the Hamiltonian (here and throughout the paper we set $\hbar=1$)
\begin{eqnarray}
H &=& \sum_{i=1}^2 \int_0^L dx \, \left(\frac{1}{2m} \partial_x \Psi_i^{\dag}\partial_x \Psi_i
+\frac{g}{2} \Psi_i^{\dag}\Psi_i^{\dag}\Psi_i\Psi_i \right)
\nonumber\\
&& -J\int_0^L dx \, \left(\Psi_1^{\dag}\Psi_2+\Psi_2^{\dag}\Psi_1\right), \label{eq:nuh}
\end{eqnarray}
where $i=1,2$ denotes one of the two wires, $g$ denotes the interaction parameter, $m$ is the mass of the atoms, $L$ is the system size, and $J$ is the tunneling matrix element between the two wires.  All the parameters are positive, and for simplicity we will use periodic boundary conditions.  The operator $\Psi_i(x)$ annihilates an
atom located at $x$ on wire $i$. We assume that initially one of the wires is occupied by atoms with the density $\rho_0$ while the other wire is initially unoccupied. At $t=0$ we couple the wires with finite, but small $J$ so that we mimic the conditions that give rise to the MQST in the two well case.  The initial chemical potential in the full wire $\mu=g\rho_0$ is simply the interaction energy per atom. Like in the two well case, the parameter
$\lambda=J/g\rho_0=J/\mu$, which is the ratio of interaction and tunneling energies, controls the transition between MQST and no MQST.  There is also a natural length scale in this system equal to the healing length of the full wire $\xi=1/\sqrt{mg\rho_0}=1/\sqrt{m\mu}$. Similarly the chemical potential $\mu=g\rho_0$ defines the (inverse) natural time scale in the system. In the natural units where $\tau=g\rho_0 t$ and $x$ is given in terms of $\xi$, the speed of sound in the full wire is $v_s=1$.

The static and dynamical properties of the system are completely governed by the following three parameters:  $\ell=L/\xi$, $K=\pi \rho_0 \xi$, and $\lambda$. In the weakly interacting regime the parameter $K\gg 1$ is equal to the effective Luttinger liquid parameter for the full wire~\cite{giamarchi_book}. In equilibrium $K$ sets the scale for quantum fluctuations in the system. As interactions increase the expression $K=\pi \rho_0 \xi$ for the Luttinger liquid parameter is no longer valid and should be substituted by another function of $\rho_0\xi$, which can be extracted from the Bethe ansatz solution~\cite{cazalilla2004}. In the regime of infinite interactions (Tonks gas) the Luttinger liquid parameter saturates at $K=1$. In this work we will focus on the regime of weak quantum fluctuations ($K/\pi \gtrsim 10$) corresponding to the semiclassical limit. As we will show in this (weakly interacting) regime the dynamics is very nontrivial. Our goal is then to treat quantum fluctuations perturbatively and analyze how they destroy the purely classical MQST. In this limit one can study the Hamiltonian (~\ref{eq:nuh}) both analytically and numerically using the semiclassical truncated Wigner method (TWA)~\cite{walls-milburn, gardiner-zoller, steel, polkovnikov2003, blakie_08, polkovnikov2009}. This method gives the leading quantum corrections to the classical (Gross-Pitaevskii) dynamics accounting for the zero point fluctuations of the bosonic creation and annihilation fields in the initial state. The inverse of the Luttinger liquid parameter plays the role of the effective Planck's constant. It can be shown that the higher order quantum corrections to TWA appearing in the form of quantum jumps are down by a factor of $1/K^2$~\cite{polkovnikov2009}. In the regime of interest such corrections are negligible so TWA is expected to give accurate quantitative description of quantum dynamics in our system.

We note that in an infinite system for a 1D gas there is no true long range order and no true condensate. At best
we have a quasicondensate at zero temperature, so we have a power law decay of correlation functions or the off-diagonal elements of the single-particle density matrix: $\langle\Psi^\dagger(x)\Psi(0)\rangle\sim 1/|x|^{1/(2K)}$. However, at finite size systems one still has a condensate fraction since a single mode of the system is macroscopically occupied compared to all other modes. Unlike in the 3D case, the condensate fraction strongly depends on the system size. But if we confine ourselves to small enough $\ell$  and large enough $K$
so that
\begin{equation}
\frac{1}{\ell^{1/2K}} = \left(\frac{\xi}{L} \right)^{1/2K} \sim O(1). \label{eq:dep}
\end{equation}
is satisfied then we have a small condensate depletion and thus a highly coherent gas.  Physically Eq.~(\ref{eq:dep}) gives us a rough estimate of the condensate fraction, and if Eq.~(\ref{eq:dep}) is satisfied we can confidently say we have a condensate.

Provided we satisfy the condition (~\ref{eq:dep}), and we have small enough $\lambda$ (as in the bosonic Josephson junction case), we show that it is possible to achieve MQST even in the 1D case.  By using an effective model description, we explicitly find the conditions that the system must satisfy in order to obtain MQST, and we numerically confirm these results. 
We also find that MQST is generally unstable, and the system eventually (on a time scale on the order of $1/J$) decays from MQST through a dynamical instability. Nevertheless, the system exhibits characteristic MQST behavior (i.e. high frequency small amplitude tunneling of atoms between the two wires) on time scales less than $O(1/J)$.  More interestingly, with a judicious choice of the system parameters $\ell$, $\lambda$, and $K$, we can use the decay mechanism to resonantly populate characteristic momentum modes $q_c$ in the initially empty wire and thus get a very sharp momentum distribution at intermediate times.  Depending on the choice of parameters, the system may or may not thermalize.  We will show what is necessary in order to achieve this by numerically solving for the dynamics for the Hamiltonian (see Eq.~(\ref{eq:nuh})).  Through our effective model, we can also attain a quantitative understanding of the underlying mechanism leading to the resonant transfer of atoms to characteristic modes and subsequent decay of MQST. 

The paper is organized as follows.  We first discuss the approximate effective description of
the system dynamics in Section~\ref{sec:analysis}. This description gives us the mechanism for the dynamical instability and a quantitative understanding of how to achieve the sharp momentum distribution at intermediate times. In Section~\ref{sec:numerics} we show specific results from the numerical solution of the
microscopic Hamiltonian~(\ref{eq:nuh}), detailing what is necessary for MQST, the question of MQST stability and the formation of the intermediate sharp momentum distribution state, and address the question of thermalization of the system at later times.  Finally in Section~\ref{sec:summary} we discuss and give an interpretation of our results.

\section{Effective Model \label{sec:analysis}}
Given the microscopic Hamiltonian in Eq.~(\ref{eq:nuh}),
consider initial conditions where the two wires are initially decoupled and all atoms occupy the
first wire ($i=1$).  Also assume that the first wire is prepared in such a way that it is close
to the ground state ($T \simeq 0$), and take the Luttinger $K$ and system size $L$
such that the depletion from the zero momentum mode is small, i.e. assume that the condition (\ref{eq:dep}) is fulfilled. Then in general for short times ($t \lesssim 1/J$) after the quench, the dynamics will be dominated by
zero momentum modes as it is in the two well systems where MQST was previously studied.

As explained in the introduction, this occurs because of the large separation of scales
present in our current system if we choose system parameters such that we are in the
MQST regime. In particular, we anticipate that the zero momentum (condensate) mode will undergo small amplitude high frequency oscillations between the two wires. We are then justified in treating the high frequency dynamics separately from long time dynamics governing non condensed modes. Thus in the leading order we can treat 1D systems as zero dimensional system and analyze condensate dynamics separately. Because of the high occupation of the zero momentum mode in the full wire, we have to use a full nonlinear treatment of the dynamics.  This can be easily done in this case, however, as we only have to treat a single mode. This mode can be treated classically (using Gross-Pitaevskii approach). In the next order of approximation we will treat this fast condensate dynamics as an external source for the non-condensate modes. We can anticipate that the momentum modes, which are resonant with the fast condensate oscillations should be excited. As we will show such process can be viewed as enhancement of the zero point fluctuations of high-momentum modes. Therefore we have to treat such modes quantum-mechanically. At short times while the occupation of the non-condensate modes is small we can linearize the resulting equations of motion (with time-dependent parameters) and obtain analytic results. At longer times excitations of these modes should provide damping to the condensate oscillations and eventual destruction of the condensate dynamics. This is a nonlinear effect happening at longer time scales, associated with the tunneling coupling $1/J$ (as opposed to the period of fast oscillations $1/\mu$). In the MQST regime we have a large separation of scales because the ratio $\lambda=J/\mu$ is small. So our strategy to address the problem is justified.

\subsection{Effective Hamiltonian for zero momentum modes}

Our starting point in the analysis of this system is to take the Hamiltonian~(\ref{eq:nuh}) and ignore terms containing nonzero momentum modes.  We then have
\begin{eqnarray}
H_0 &=& -J(\Psi_{1,0}^{\dag} \Psi_{2,0}+\textrm{h.c.}) + \frac{g}{2L}\sum_{i=1}^2
\Psi_{i,0}^{\dag}\Psi_{i,0}^{\dag}\Psi_{i,0}\Psi_{i,0},
\nonumber\\
&&\label{eq:nuh0}
\end{eqnarray}
where $\Psi_{i,0}$ annihilates a particle in wire $i$ with zero momentum ($q=0$).
The above Hamiltonian~(\ref{eq:nuh0}) can be mapped into a spin system via Schwinger representation:
\begin{eqnarray}
S_x &=& \frac{\Psi_{1,0}^{\dag}\Psi_{2,0}+\Psi_{2,0}^{\dag}\Psi_{1,0}}{2}  \label{eq:mapspin1}\\
S_y &=& \frac{\Psi_{1,0}^{\dag}\Psi_{2,0}-\Psi_{2,0}^{\dag}\Psi_{1,0}}{2i} \label{eq:mapspin2}\\
S_z &=& \frac{\Psi_{1,0}^{\dag}\Psi_{1,0}-\Psi_{2,0}^{\dag}\Psi_{2,0}}{2}. \label{eq:mapspin3}
\end{eqnarray}
Then the Hamiltonian $H_0$ can be written as
\begin{eqnarray}
H_0 &=& -2JS_x+\frac{g}{L}S_z^2. \label{eq:nuh0spin}
\end{eqnarray}
For a system of $N$ atoms, Eqs.~(\ref{eq:mapspin1})-(\ref{eq:mapspin3}) map the bosons to the spin with the magnitude $S=N/2$. We can easily find the Heisenberg equations of motion from the $SU(2)$ algebra. Since we pick $K$ large enough to be in the semiclassical limit and the zero momentum mode is highly occupied initially, we can safely ignore the effects of quantum fluctuations and treat $H_0$ classically. Furthermore, since depletion is small, we can take $N$ to be the total number of atoms in the system. Using energy and total spin conservation (particle number conservation in original fields), we can write down the equation of motion for $S_z$
\begin{eqnarray}
\partial_t^2 S_z &=& -\left(4J^2-\frac{2gE_0}{L}\right)S_z-\frac{2g^2}{L^2}S_z^3,
\end{eqnarray}
where $E_0$ is the total (conserved) energy of the system. Initially all atoms are in the first wire, so we have $S_x=S_y=0$ and $S_z=N/2$ as our initial conditions, which uniquely determine $E_0$. It is convenient to work with  the dimensionless time $\tau=g\rho_0 t=gNt/L$ and rescale the spin $S_z\to n=S_z/S=(N_1-N_2)/N$ so that it represents the scaled population difference between the two wires. Then the equation of motion reads (see also Ref.~\cite{polkovnikov2002})
\begin{eqnarray}
\partial_\tau^2 n &=& -\left(4\lambda^2-\frac{1}{2}\right)n-\frac{1}{2}n^3, \label{eq:n}
\end{eqnarray}
where $\lambda=J/g\rho_0=J/\mu$. The parameter $\lambda$ determines whether or not we have
MQST.  Eq.~(\ref{eq:n}) is equivalent to the equation of motion for a single particle in the
potential
\begin{eqnarray}
V(n)&=& \left(4\lambda^2-\frac{1}{2}\right)\frac{n^2}{2}+\frac{n^4}{8}
\end{eqnarray}
and total energy $E=(\partial_{\tau}n)^2/2+V(n)$.  For $\lambda>1/\sqrt{8}$ the potential
$V(n)$ has a single minimum at $n=0$ (equal population in both wires), while for
$\lambda<1/\sqrt{8}$ the minimum at $n=0$ becomes a local maximum and two new minima appear at
\begin{equation}
\pm n_{\star},\quad{\rm where}\;n_\star=\sqrt{1-8\lambda^2}. \label{eq:nstar}
\end{equation}
Nonzero $n$ signifies population imbalance, so a necessary condition for MQST is
$\lambda<1/\sqrt{8}$.  But this condition is not sufficient for MQST, since a system with
high enough energy $E>V(0)=0$ will overcome the potential barrier and not self-trap.
We also need $E<0$.  For our initial conditions of complete imbalance ($n(0)=1,\; \partial_t n(0)=0$) we have $E=2\lambda^2-1/8$, so in order to have MQST we need to have
\begin{equation}
\lambda<1/4.
\end{equation}

\begin{figure}
	\includegraphics[scale=0.7,angle=270]{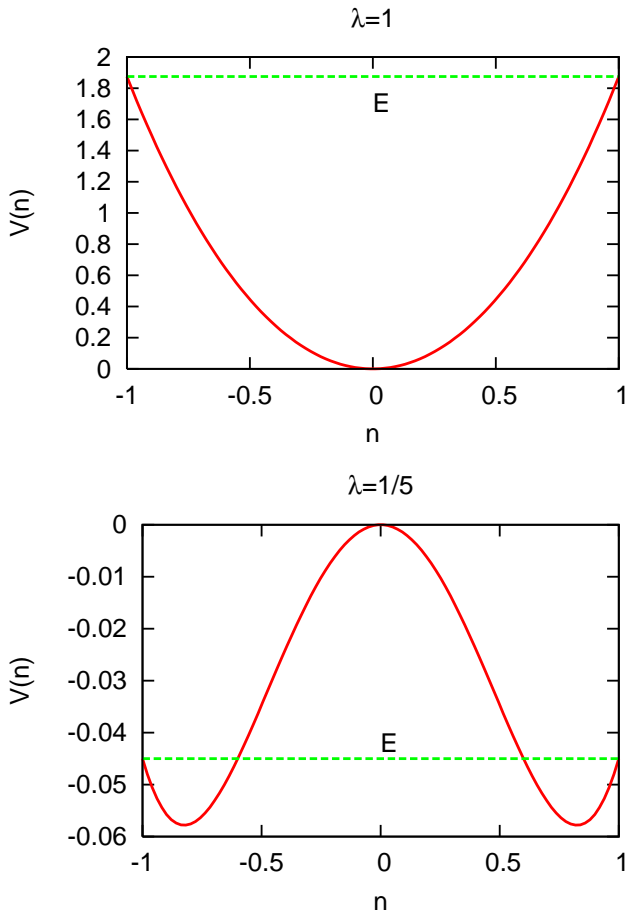}
		\caption{(Color online) Plot of $V(n)$ for $\lambda=1$ (top, not self trapped)
and $\lambda=1/5$ (bottom, self trapped),
along with the energies $E(\lambda)$ for both cases. Note that we have the constraint
$-1<n<1$ from total number conservation.
 \label{fig:zeromode} }
\end{figure}

In Figure~\ref{fig:zeromode} we show the effective single particle potential, along with
the energies, for $\lambda=1$ where there is no self trapping, and $\lambda=1/5$ where we
have MQST.  We can write down the exact solution for Eq.~(\ref{eq:n}) in terms of a Jacobi
elliptic function
\begin{equation}
n(\tau)=\textrm{cn} \left(2\lambda \tau,\frac{1}{4\lambda}\right). \label{eq:soln}
\end{equation}
Since we are interested in the self trapped regime, where $\lambda$ is small, we can expand
Eq.~(\ref{eq:soln}) to second order in $\lambda$ and ignore higher order terms
\begin{equation}
n(\tau) \simeq 1-8\lambda^2 \sin^2 \frac{n_{\star}\tau}{2}, \label{eq:solnap}
\end{equation}
where $n_{\star}$ is given by Eq.~(\ref{eq:nstar}). This approximation is very good for
$\lambda$ as high as $1/5$ (recall that we need $\lambda<1/4$ in order to have MQST). In MQST we only have a partial tunneling of atoms between the two wires.  Atoms oscillate back and forth between the wires, with a frequency $n_{\star}\simeq 1$ in dimensionless units (which corresponds to the frequency $\mu$ in original units) and the amplitude $4\lambda^2$.  So in MQST we have high frequency and small amplitude transfer of atoms between both wires. Whenever the effect of quantum fluctuations is not important, this situation persists and MQST
is stable. As we described above these condensate oscillations couple to the non-condensate modes. Our next step will be to analyze the effect of this coupling on dynamics.

\subsection{Effective Hamiltonian for nonzero momentum modes}

We will now focus on finding the effective Hamiltonian for the nonzero momentum modes.
We will employ the following strategy, as explained in the introduction of this section.
We go back to the microscopic Hamiltonian and treat the
zero momentum mode fields as \emph{classical time dependent sources} coupled to nonzero momentum modes,
their time dependent behavior given by the solution to the equations for zero momentum mode fields
described above.  We then keep contributions from nonzero
momentum terms up to quadratic order, and solve the equations of motion for all nonzero momentum modes.
The dynamics described by this effective Hamiltonian
then should be valid as long as the occupation in the nonzero momentum modes remains small.  Through this
effective Hamiltonian, we will find the mechanism of MQST decay and illustrate the importance of quantum fluctuations.  We will then need to use the explicit solution for the zero momentum mode fields
$\Psi_{1,0}(\tau)$ and $\Psi_{2,0}(\tau)$, which we can easily find from solving the classical
system given by Eq.~(\ref{eq:nuh0spin}).
For $\lambda<1/5$ to a very good approximation  (up to terms of the order of  $\lambda^4$) the solution is
\begin{eqnarray}
&&\Psi_{1,0}(\tau) = \sqrt{N}\sqrt{1-4\lambda^2 \sin^2\frac{n_{\star} \tau}{2}}\nonumber\\
&&~~\times\exp\left[i\left(\phi_{1,0}-\tau+ \lambda^2 \left(\tau-\frac{\sin n_{\star}\tau}{n_{\star}}\right)\right)\right],\label{eq:psi0a}\\
&&\Psi_{2,0}(\tau) = \sqrt{N}\, 2\lambda \sin\frac{n_{\star} \tau}{2}\nonumber\\
&&~~\times\exp\left[i\left(\phi_{2,0}-\frac{\tau}{2}-\lambda^2 \left(\tau-\frac{\sin n_{\star}\tau}{n_{\star}}\right)\right)\right], \label{eq:psi0}
\end{eqnarray}
where $\phi_{i,0}$ is the initial phase in the zero momentum mode for wire $i$ and $N$ is the total number
of atoms. Since the our initial state is the ground state of \emph{decoupled wires}, these phases are random and in order to compute the expectation value of an arbitrary operator one needs to take average over them (technically this follows from the fact that in the classical limit the Wigner function representing initial Fock state reduces to the fixed amplitude of $\Psi_{1,2}$ and the random phase~\cite{polkovnikov2002, polkovnikov2003a}). But in our particular case these phases are not important because initially the second wire is empty so one can safely set them to zero. Note Eqs.~(\ref{eq:psi0a}) and (\ref{eq:psi0}) have a running phase which is typical of states in MQST. Also in the range of $\lambda<1/5$ we are interested in we can set $n_{\star}=1$ to a very good approximation.

Next we will expand the Hamiltonian to the quadratic order in non-zero momentum modes. Doing this we will also ignore contributions to the Hamiltonian of order $\lambda^2$ and higher. This approximation is justified for small $\lambda$, and small enough system size such that $\ell \lambda=JL/\mu\xi\lesssim 1$, where $\xi$ is the coherence length in the full wire. The constraint on system size is due to the fact that we need characteristic kinetic energies in the system ($\propto 1/\ell^2$) to be larger than the terms of order $\lambda^2$ ignored in the Hamiltonian. This requirement on the system size not only simplifies the analysis, but (as we show later) is also crucial in order to obtain a sharp momentum distribution at intermediate times ($\tau \sim 1/\lambda$).

The effective Hamiltonian for the nonzero momentum modes can be written as $H=\sum_{q>0}H_q$. As before, we will employ dimensionless units (or equivalently, all energies in terms of $\mu$, the natural inverse time unit). There are three contributions to $H_q$, which we will consider and discuss separately:
\begin{eqnarray}
H_q &=& H_q^{F}+H_q^{E}+H_q^J,
\end{eqnarray}
where $H_q^{F}$ is the contribution from the full wire, $H_q^{E}$ is from the empty wire, and
$H_q^J$ is the contribution from tunneling between the wires.
For the full wire, we have
\begin{eqnarray}
H_q^F &=& \Psi_{1q}\Psi_{1,-q}e^{2i\tau}+\Psi_{1q}^{\dag}\Psi_{1,-q}^{\dag}e^{-2i\tau} \nonumber\\
&&+(2+T_q)(\Psi_{1q}^{\dag}\Psi_{1q}+\Psi_{1,-q}^{\dag}\Psi_{1,-q}),
\label{eq:LL}
\end{eqnarray}
where $T_q=q^2/2m\mu$, the scaled kinetic energy for momentum $q$.
Note that we have an explicit time dependence in this contribution coming from the running phase.
The phase monotonically increases at a rate $1$ in dimensionless time (with rate $\mu$ in original
units). This is also the frequency with which the atoms tunnel between wires in the self trapped
regime.  Physically, this contribution couples an external harmonic source with a frequency $\mu$
to creation and annihilation operators for particle pairs with momenta $\pm q$.  This acts like
an external energy pump, and is the cause of the dynamical instability that develops in the
system due to resonance of the external frequency $\mu$ with the natural system modes.
For the empty wire, we have
\begin{eqnarray}
H_q^E &=& T_q (\Psi_{2q}^{\dag}\Psi_{2q}+\Psi_{2,-q}^{\dag}\Psi_{2,-q}). \label{eq:FB}
\end{eqnarray}
Note that no explicit time dependence shows up in this contribution if we ignore terms of order
$\lambda^2$, and that Eq.~(\ref{eq:FB}) is just the Hamiltonian for free Bosons of momentum $q$.
There is a subtle issue in 1D, where in the ground state of a dilute interacting Bosonic gas the interaction energy is always higher than the kinetic energy, since the interaction energy is proportional to density while kinetic energy is proportional to density squared.  At short times the density in the empty wire is very small, so naively it seems that we wrongfully ignored the interaction.  However, this consideration is somewhat misleading for two reasons. First, we have included effects of interactions in the empty wire already: recall that we treated the zero momentum modes separately, and in this treatment we have included the effect of interaction in the zero momentum modes. As long as depletion from the quasicondensate is small, which it is for cases we consider, the interaction terms involving nonzero momentum modes are negligible compared to their kinetic energy. Second, we are dealing with far from equilibrium systems and the momentum modes which are populated have very high energy of the order of the chemical potential in the full wire. While the particle density in the initially empty wire remains small the interaction energy is negligible compared to the kinetic energy and can be safely neglected. Finally we are left with the contribution for the coupling between the two wires, which
is simply
\begin{eqnarray}
H_q^J &=& -\lambda(\Psi_{1q}^{\dag}\Psi_{2q} + \Psi_{1,-q}^{\dag}\Psi_{2,-q} +\textrm{h.c.}).
\end{eqnarray}

The explicit time dependence of $H_q$ can be eliminated by a simple unitary transformation:
\begin{equation}
\tilde H_q=\mathcal{U}_q H_q \mathcal{U}_q^{\dag} +i\left(\partial_{\tau}\mathcal{U}_q\right)
\mathcal{U}_q^{\dag},
\label{tildeH}
\end{equation}
where
\begin{equation}
\mathcal{U}_q = \exp\left(i \tau\sum_{i=1}^2(\Psi_{i,q}^{\dag}\Psi_{i,q}+\Psi_{i,-q}^{\dag}
\Psi_{i,-q}) \right),
\end{equation}
where the second term in (\ref{tildeH}) reflects the fact that $\mathcal{U}_q$ is time dependent.
Physically this comes about because we are going to an accelerating (rotating) frame. The new Hamiltonian $\tilde H_q$ no longer has any explicit time dependence.

We then have for each of the contributions in turn
\begin{eqnarray}
\tilde H_q^{F} &=& \Psi_{1q}\Psi_{1,-q}+\Psi_{1q}^{\dag}\Psi_{1,-q}^{\dag}\nonumber\\
       &&+(1+T_q)(\Psi_{1q}^{\dag}\Psi_{1q}+\Psi_{1,-q}^{\dag}\Psi_{1,-q})\label{eq:rotea}\\
\tilde H_q^{E} &=&  (T_q-1) (\Psi_{2q}^{\dag}\Psi_{2q}+\Psi_{2-q}^{\dag}\Psi_{2-q})\label{eq:roteb}\\
\tilde H_q^{J} &=& -\lambda(\Psi_{1q}^{\dag}\Psi_{2q} + \Psi_{1-q}^{\dag}\Psi_{2-q} +\textrm{h.c.}).
\label{eq:rote}
\end{eqnarray}
Note that the unitary (gauge) transormation simply results in $T_q \rightarrow T_q-1$ i.e. subtracting the chemical potential of the full wire from the kinetic energy.  The difference in sign in the term $T_q\pm 1$ between the full and empty wire has a simple physical interpretation:  in order for a particle to tunnel from the empty wire to the full wire it must overcome the difference in interaction energies in the wire. In other words, the chemical potential difference between the two wires acts as a bias (uniform) potential, with the full wire at a higher potential due to its interaction energy.

Since $\tilde H_q$ is quadratic, we can explicitly diagonalize it. Instead of bosonic fields $\Psi_{i,q}$ and $\Psi^\dagger_{i,q}$ it is convenient to work with the fields which represent density and phase fluctuations:
\begin{eqnarray}
\Psi_{i,q} &=& \frac{n_{i,q}+i\phi_{i,q}}{\sqrt{2}} \nonumber\\
\Psi_{i,-q}^{\dag} &=& \frac{n_{i,q}-i\phi_{i,q}}{\sqrt{2}}.
\end{eqnarray}
Since this is a canonical transformation the new fields satisfy standard coordinate-momentum like commutation relations:
\begin{eqnarray}
&&\lbrack n_{i,q}, \phi_{j,q'} \rbrack= i\delta_{q,-q'}\delta_{i,j} \nonumber\\
&&n_{i,q}^{\dag} = n_{i,-q},\quad \phi_{i,q}^{\dag} = \phi_{i,-q}. \label{eq:nprop}
\end{eqnarray}
Their Fourier transforms $n_i(x)=1/L \sum_q n_{i,q} \exp(iqx)$ and
$\phi_i(x)=\sum_q \phi_{i,q} \exp(iqx)$ satisfy similar commutation relations in real space
\begin{eqnarray}
&&\lbrack n_i(x),\phi_i(x') \rbrack = i\delta(x-x')\delta_{i,j} \nonumber\\
&&n_i(x)^{\dag} = n_i(x) \quad \phi_i(x)^{\dag} = \phi_i(x) \label{eq:npropx}.
\end{eqnarray}
which are the correct relation for physical number and phase operators.

To see that the fields $n_{1,q}$ and $\phi_{1,q}$ indeed correspond to the correct number and phase operators for small depletion we observe that the Fourier component of the density
\begin{eqnarray}
&&\sum_{q'} \Psi_{1,q'}^{\dag}\Psi_{1,q+q'} \simeq \Psi_{1,0}^{\dag}\Psi_{1,q} + \Psi_{1,-q}^{\dag}\Psi_{1,0} \nonumber\\
&&\simeq \sqrt{N}\left(\Psi_{1,q}+\Psi_{1,-q}^{\dag} \right)= \sqrt{2N} n_{1,q}.
\end{eqnarray}
is exactly proportional to $n_{1,q}$. Since $\phi_{1,q}$ represents the conjugate field to $n_{1,-q}$, then up to a constant factor ($1/\sqrt{2N}$) it is just the Fourier component of the physical phase (this can be checked by explicit calculation). The approximate equality above is only valid when the depletion is small and the terms with $q'=0,-q$ dominate the sum. Note that this argument cannot be applied to $n_{2q}$ and $\phi_{2q}$, the fields in the second wire, since we have a small occupation in the zero momentum mode. $n_{2q}$ and $\phi_{2q}$ therefore do not correspond to physical number and phase fluctuations in the empty wire, but for simplicity we will use the same terminology in referring to them, since their properties are otherwise identical to the number and phase operators in the full wire.

We can then write our Hamiltonian as
$\tilde H_q=\tilde H_q^F+\tilde H_q^E+\tilde H_q^J$ where
\begin{eqnarray}
\tilde H_q^{F} &=& T_q\phi_{1q}\phi_{1,-q}+(T_q+2)n_{1q}n_{1,-q} \label{eq:hqprimea}\\
\tilde H_q^{E} &=& (T_q-1)\left(n_{2q}n_{2,-q}+\phi_{2q}\phi_{2,-q}\right)  \label{eq:hqprimeb}\\
\tilde H_q^{J} &=& -\lambda\left(n_{1q}n_{2,-q}+n_{2q}n_{1,-q} \right. \nonumber\\
&& \left.+\phi_{1q}\phi_{2,-q}+\phi_{2q}\phi_{1,-q}\right) \label{eq:hqprime},
\end{eqnarray}
We can easily find the expectation values of the occupation of momentum modes by real particles in the full and empty wires using the following relation
\begin{eqnarray}
\Psi_{i,q}^{\dag}\Psi_{i,q}(t) &=& \frac{1}{2}\left( |n_{i,q}(t)|^2 + |\phi_{i,q}(t)|^2 \right).
\end{eqnarray}

\subsection{Dynamics of nonzero momentum modes}

The Hamiltonian $\tilde H_q$ in Eq.~(\ref{eq:hqprime}) is quadratic and therefore it is easy to analyze the dynamics in our system.  Before we go on to solve the system $\tilde H_q$ exactly let us gain some physical insight by doing a perturbative calculation for the system treating $\tilde H_q^{J}$
as the perturbation. In the empty wire the Hamiltonian $\tilde H_q^{E}$ is diagonalized in
the original particle basis (see Eq.~(\ref{eq:roteb})). Its spectrum is $\omega_E(q)=T_q-1$ (in the original units $\omega_E(q) = q^2/(2m)-\mu$). Apart from subtraction of the chemical potential due to our gauge transformation, this is just the energy spectrum for free bosons.

Next, let us diagonalize the Hamiltonian for the full wire $\tilde H_q^{F}$ (\ref{eq:hqprimea}). This can be done in the quasiparticle basis $\psi_q$
\begin{eqnarray}
\phi_{1q} &=& \left(1+\frac{2}{T_q}\right)^{1/4}  \frac{i\psi_{-q}^{\dag}-i\psi_q}
{\sqrt{2}}\\
n_{1q}    &=& \left(1+\frac{2}{T_q}\right)^{-1/4} \frac{\psi_{-q}^{\dag} + \psi_q}{\sqrt{2}},
\end{eqnarray}
where $\psi_q$ and $\psi_q^{\dag}$ obey the usual commutation relations
$\lbrack \psi_q , \psi_{q'}^{\dag} \rbrack = \delta_{qq'}$.
In terms of quasiparticles $\tilde H_q^F$ is given by
\begin{eqnarray}
\tilde H_q^{F} &=& \sqrt{T_q(2+T_q)}\left(\psi_q^{\dag}\psi_q + \psi_{-q}^{\dag}\psi_{-q}+1 \right).
\end{eqnarray}
The energy spectrum for the full wire $\omega_F(q)=\sqrt{T_q(2+T_q)}$ is just the usual
Bogoliubov spectrum for Bose gases close to the ground state. In the original energy units this spectrum reads
\begin{equation}
\omega_F(q) = \sqrt{\frac{q^2}{2m}\left(2\mu+\frac{q^2}{2m}\right)}= v_s|q| \sqrt{1+\frac{q^2}{(2mv_s)^2}}, \label{eq:omf}
\end{equation}
where $v_s=\sqrt{\mu/m}$ is the speed of sound in the full wire.  For small momenta
$\omega_F(q) \simeq v_s|q|$ and we have the usual sound modes for
a Bose gas close to zero temperature.  Note however that we start in a high energy state
with respect to the coupled Hamiltonian (interaction energy $\sim \mu$),
and we need to take into account the correct spectrum for high momentum modes $q\sim mv_s$
as given by Eq.~(\label{eq:omf}).

In order to treat $\tilde H_q^{J}$ perturbatively, we will write it in the $\psi,\Psi_2$ basis,
which diagonalizes $\tilde H_q^{F}$ and $\tilde H_q^{E}$ respectively. In this basis,
$\tilde H_q^{J}$ has terms with all possible quadratic combinations of $\psi,\Psi_2,\psi^{\dag},
\Psi_2^{\dag}$.  Consider that our initial state has very small depletion from the zero momentum modes,
and therefore all contributions of $\tilde H_q^J$ with at least one annihilation operator have
vanishingly small matrix elements compared with contributions of $\tilde H_q^J$ with two creation
operators.  So for our perturbative calculation to a very good approximation
\begin{equation}
\tilde H_q^{J} \approx \frac{\lambda}{2} \left[\frac{\displaystyle{\sqrt{2+T_q}-\sqrt{T_q}}}
{\displaystyle{\left(T_q(2+T_q)\right)^{1/4}}}\right]
(\psi_{-q}^{\dag}\Psi_{2,q}^{\dag}+\psi_q^{\dag}\Psi_{2,-q}^{\dag}) \label{eq:vq},
\end{equation}
It is clear from Eq.~(\ref{eq:vq}) that the physical process that drives the depletion from the
zero momentum modes and destroys MQST is the production of quasiparticle pairs with opposite momenta $\pm q$ in each wire.  At long times the dominant contribution to particle creation will come from the momentum mode satisfying the resonant condition
\begin{equation}
\omega_E(q_c)+\omega_F(q_c)=(T_{q_c}-1)+\sqrt{T_{q_c}(2+T_{q_c})}=0 \label{eq:charq}
\end{equation}
which has the solution
\begin{equation}
T_{q_c}=\frac{q_c^2}{2m\mu}=\frac{1}{4}. \label{eq:Tqc}
\end{equation}
In the original units Eq.~(\ref{eq:charq}) reads
\begin{equation}
\left(\frac{q_c^2}{2m}-\mu\right) + v_s|q_c|\sqrt{1+\frac{q_c^2}{(2mv_s)^2}}=0, \label{eq:charqu}
\end{equation}
with the solution
\begin{equation}
q_c=\sqrt{\frac{m\mu}{2}}=\frac{mv_s}{\sqrt{2}}, \label{eq:qc}
\end{equation}
This wavevector describes the characteristic momentum where the zero momentum mode particles scatter with the highest probability, possibly causing a breakdown of MQST. As we will see later $q_c$ represents the momentum of the maximally unstable mode.

Note that Eq.~(\ref{eq:charqu}) can be rewritten in the following way, giving a clear physical interpretation:
\begin{equation}
\frac{q^2}{2m} + v_s|q|\sqrt{1+\frac{q^2}{(2mv_s)^2}}=\mu. \label{eq:charqs}
\end{equation}
The right hand side of Eq.~(\ref{eq:charqs}) is just the sum of the energies of two
quasiparticles of momentum $q$, with one quasiparticle in the full wire and the other in the empty wire, while the left hand side is just the frequency of oscillations of the zero momentum mode, which serves the external source term. Through the resonant coupling particles are scattered out of the zero-momentum mode to the momenta close to $q_c$. Because of the bosonic nature of quasi-particle, we find that the population of the mode with $q=q_c$ increases exponentially with the rate $2J/\sqrt{3}$.  If the number of unstable modes is small we can expect a very sharp momentum distribution forming in both empty and full wires at time scales of the order $1/J$.

Now we outline the exact solution for the dynamics of the system in the linearized regime.
We can easily find the Heisenberg equations of motion for the nonzero momentum modes using
Eqs.~(\ref{eq:hqprimea})-(\ref{eq:hqprime})
\begin{eqnarray}
\partial_{\tau}n_{1q} &=& T_q\phi_{1q}-\lambda \phi_{2q}\\
\partial_{\tau}\phi_{1q} &=& -(T_q+2)n_{1q}+\lambda n_{2q}\\
\partial_{\tau}n_{2q} &=& (T_q-1)\phi_{2q}-\lambda \phi_{1q}\\
\partial_{\tau}\phi_{2q} &=& -(T_q-1)n_{2q}+\lambda n_{1q}. \label{eq:eom}
\end{eqnarray}
Assuming the solutions have a time dependence $\sim \mathrm e^{-i\omega(q) \tau}$,
we find two pairs of eigenfrequencies for each momentum: $\pm\omega_\pm(q)$ (measured in units of $\mu$):
\begin{equation}
\omega_{\pm}^2(q) = T_q^2+\lambda^2+\frac{1}{2} \pm \frac{1}{2}\sqrt{(4T_q-1)^2+4\lambda^2
(4T_q^2-1)}.\label{eq:omega}
\end{equation}
Note that in the self trapped regime ($0<\lambda<1/4$) the term under the radical in Eq.~(\ref{eq:omega}) can be negative. Thus, there is always an interval in $T_q$ where $\omega_{\pm}$ is complex.

\begin{figure}[pt]
	\centering
		\includegraphics[scale=0.7,angle=270]{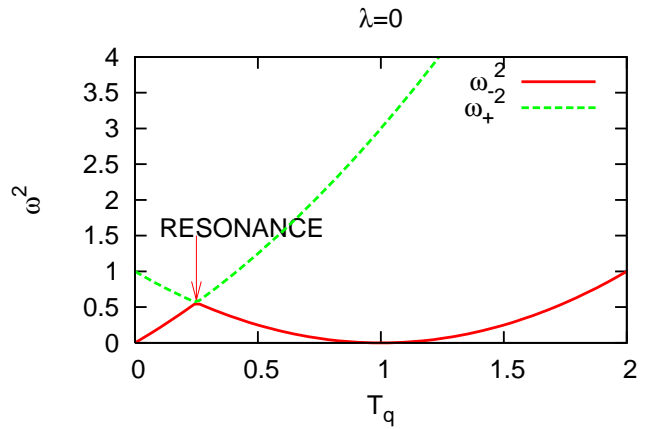}
		\caption{(Color online) Plot of $\omega_{\pm}^2$ as a function of $T_q$ at $\lambda=0$.
Solid red line corresponds to $\omega_-^2$ and
dashed green line corresponds to $\omega_+^2$. Note that the two curves cross at $T_q=1/4$,
which is the resonance point (see Eq.~(\ref{eq:Tqc})) where the
frequency of oscillations in the zero momentum modes matches the energy required to create
particle pairs of characteristic momenta $\pm q_c$ (see Eq.~(\ref{eq:qc}))
in each wire . \label{fig:lambzer} }
\end{figure}

\begin{figure}[pt]
	\centering
		\includegraphics[scale=0.7,angle=270]{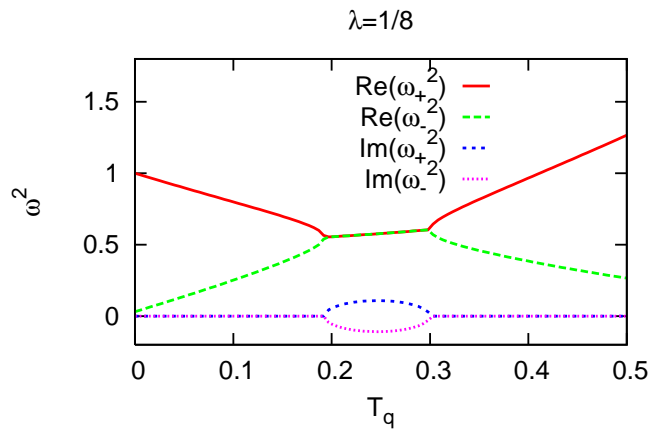}
		\caption{(Color online) Same as in Fig.~\ref{fig:lambzer} but for $\lambda=1/8$.
In this case $\omega_{\pm}$ acquires an imaginary part in the region around the resonance point (see Eq.~(\ref{eq:Tqc})), so we have a dynamically unstable set of momenta in this region.  The
extent of the unstable region is $\sqrt{3}\lambda/2$ while the maximum peak for the
imaginary part of $\omega_{\pm}$ is $\lambda/\sqrt{3}$.  The momentum
modes falling within the unstable region then become populated at an exponential rate,
in a time scale $\sim 1/\lambda$ in dimensionless units ($1/J$ in
original units). \label{fig:lambfin} }
\end{figure}

In order to get a better physical understanding of the expression in Eq.~(\ref{eq:omega})
let us first analyze the trivial limit $\lambda=0$ where Eq.~(\ref{eq:omega}) reduces to
\begin{equation}
\omega_{\pm}^2(q) = T_q^2+\frac{1}{2} \pm \frac{1}{2}|4T_q-1|. \label{eq:lambzer}
\end{equation}
The two branches of the spectrum $\omega_{\pm}^2(q)$ are plotted in Figure~\ref{fig:lambzer}.  The two curves cross at $T_q=1/4$, the resonance point we found in the perturbative calculation (see Eq.~\ref{eq:Tqc}). The interpretation of $\omega_{\pm}$ at $\lambda=0$ here is simple. For $0<T_q<1/4$, $\omega_{+}^2=(T_q-1)^2$ corresponds to the spectrum of the free Bose gas in the empty wire shifted to the right by an amount equal to the chemical potential difference between the wires.  Likewise, for $0<T_q<1/4$, $\omega_{-}$ is the spectrum of the
interacting Bose gas in the full wire, and for $T_q>1/4$, $\omega_+$ and $\omega_-$ switch roles.
For $\lambda=0$, the two wires do not interact with each other and their spectra are
completely independent.  For small nonzero $\lambda$, however, the crossing point in Figure~\ref{fig:lambzer} gives the resonance condition we found from the perturbative calculation.  It is easy to see that this is the channel for MQST decay.

Next we take $\lambda$ small but nonzero.  We plot $\omega_{\pm}$ for $\lambda=1/8$ in Figure~\ref{fig:lambfin}. In this case $\omega_{\pm}$ acquires an imaginary part. This signals the development of a dynamical instability in the region where $\omega_{\pm}$ is complex.  The dynamically unstable region is centered
around the resonance point found in our perturbative calculation ($T_q=1/4$),
and the extent of the region is $\simeq \sqrt{3}\lambda/2$. In terms of momentum, the dynamically
unstable region is centered around $q_c=mv_s/\sqrt{2}$ (where the speed of sound $v_s=\xi \mu$),
and the extent of the region is $\sqrt{3}\lambda q_c$.  From the exact spectrum of $\tilde H_q^{J}$ we then find that there is an unstable region of finite extent $\sim \lambda$ around the resonance point, instead
of the single characteristic mode found from the perturbative calculation.  It then follows that the population of all momentum modes that fall within the unstable region, will exponentially grow in time.  The time scale
in which this happens is given by the inverse of the typical value of the imaginary part of $\omega_{\pm}$, which is of the order $1/\lambda$ ($1/J$ in original energy units).  The maximum rate of increase possible is determined from the peak of the imaginary part of $\omega_{\pm}$, which is $2J/\sqrt{3}$. This is precisely what we obtained from the perturbative Fermi Golden Rule calculation.

\begin{figure*}[ht]
\centering
\scalebox{0.9}{\includegraphics{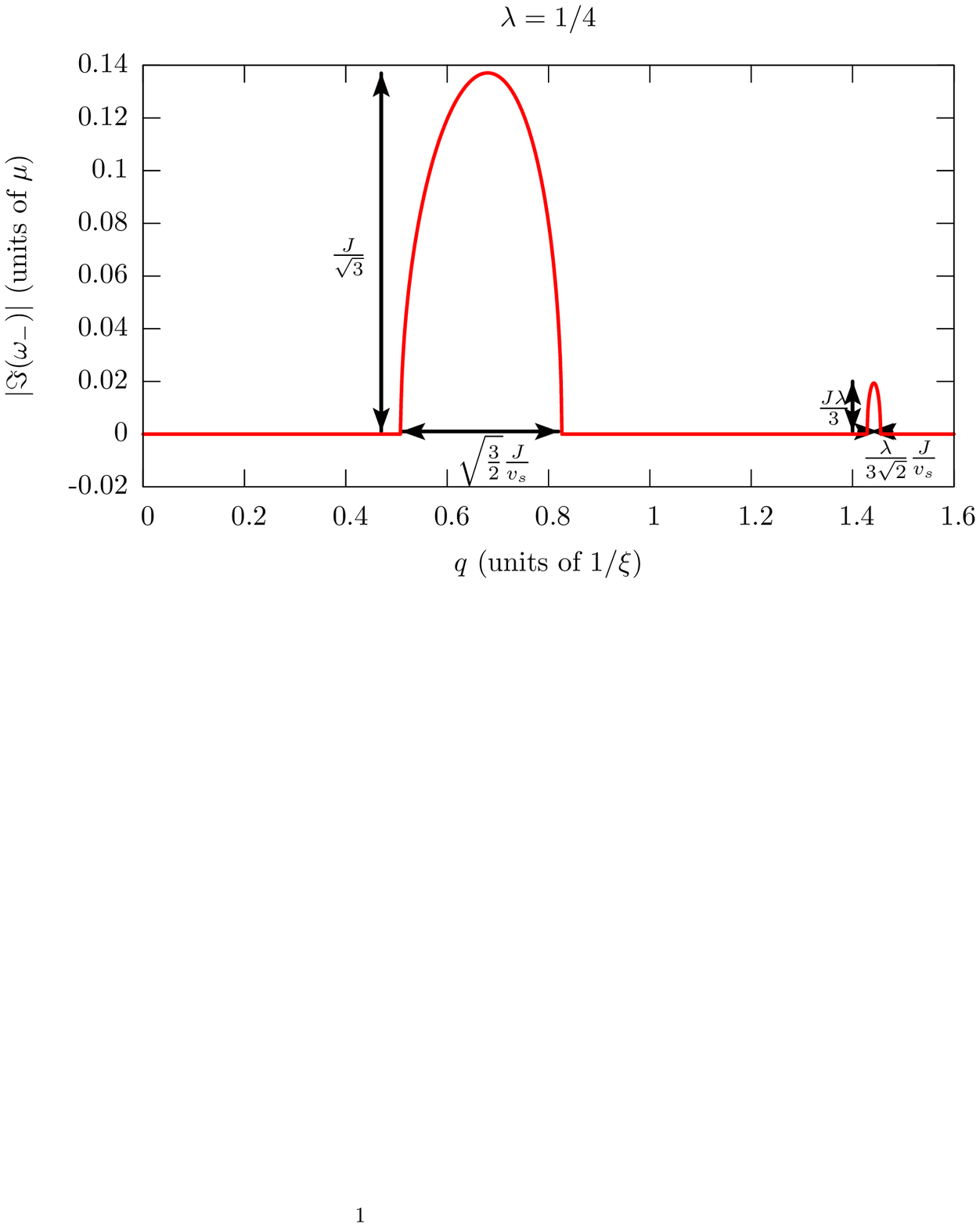}}
\caption{(Color online) Plot of $|\Im(\omega_-)|$ (in units of $\mu$) vs $q$
(in units of $1/\xi$). Unlike $\omega_+$ where there
is only one region where it acquires an imaginary part, for $\omega_-$ we have two such regions.
The first region is exactly the same as in $\omega_+$, centered around $q=q_c=1/\sqrt{2}\xi$. The second unstable region occurs only for $\omega_-$ around $q=2q_c$.  Both the extent of the region of instability and the maximum of the imaginary part of $\omega$ in the second region are of order $\lambda$ times smaller than the corresponding quantities in the first region, meaning that the instability in the second region is suppressed compared to the one in the first region if $\lambda$ is small.\label{fig:imag} }
\end{figure*}

\subsection{Second dynamically unstable region}
Let us briefly discuss other regions of interest in the system.
From Eq.~(\ref{eq:omega}) we find that there is an additional instability occurring around $2q_c$, which is twice the resonant momentum. However, this time only $\omega_-$, and not $\omega_+$, becomes complex. We then have a second region of dynamical instability in the system. It is easy to see that there are no additional unstable regions in the system. This instability emerges because $\omega_-^2(2q_c)$ becomes negative for any small positive $\lambda$. This instability, however, is much weaker since the maximum imaginary part of the frequency is $\lambda^2/3$. Since the main instability around $q_c$ develops on times of the order of $1/\lambda$ the effect of the second instability on dynamics is usually very small. Nevertheless, as we will see from full nonlinear numerical simulations, this instability is not just an artifact of the effective model and the corresponding mode is indeed excited in the system. In Fig.~\ref{fig:imag} we plot imaginary part of $\omega_-(q)$ for $\lambda$ close to $1/4$, i.e. to the onset of self-trapping. It is clear this instability also has a smaller width (by a factor $\lambda/3\sqrt{3}$) compared to the first instability. For smaller values of $\lambda$ the effect of the second instability is even weaker.

Let us briefly mention that the physical process associated with this instability can be understood by looking into perturbation theory. Since the instability here is on the order of $\lambda$ smaller than the one associated with the first unstable region, we expect the effect to show up in second order perturbation theory where we have contributions of order $\lambda^2$.  This is indeed the case, and we find there is a resonance here similar to what occurs in first order.  In second order, we have a contribution proportional to the matrix elements of $\Psi_{2,q}^{\dag}\psi_{q}\psi_{q}^{\dag}\Psi_{2,-q}^{\dag}$ and likewise for $q\rightarrow -q$. This contribution has the net effect of producing particle pairs of opposite momenta in the empty wire, where the resonance condition for these terms is now given by $2(T_q-1)=0$ instead of Eq.~(\ref{eq:charq}). The characteristic mode then satisfies $2(q^2/2m-\mu)=0$ which gives us $q=2q_c$.  This process can be interpreted as two atoms in the full wire zero momentum mode (with interaction energy $\mu$ per atom) tunneling into the empty wire, where all of their interaction energy is converted into kinetic energy.  There is a similar process in the full wire as well, where two atoms in the zero momentum mode are transfered to the $2q_c$ mode and in the process all their interaction energy is converted into kinetic energy.  One can show that to all orders of perturbation theory only $q_c$ and $2q_c$ are resonantly populated, which is consistent with the exact solution of the effective model.

\subsection{Effect of system size}
The finite system size plays a crucial role in our analysis. First, as we noted earlier, we are relying on having a large condensate fraction in the initially full wire. This is only possible in one dimension if the system size is finite and small enough (see Eq.~(\ref{eq:dep})). Additionally finite system size leads to quantization of allowed momentum modes. When we analyzed the spectrum of unstable modes above we found that there are always unstable regions corresponding to imaginary frequencies. However, having these unstable regions does not imply that a finite size system actually has unstable modes and \emph{always} decays from MQST. The dynamical instability rather provides us with a possible route of MQST decay. The more accurate statement is that weakly interacting coupled 1D systems that initially exhibit MQST will undergo MQST decay through the resonant mechanism described above as long as there are actual momentum modes in the system that fall within the dynamically unstable region. The set of allowed modes is controlled by the system size, so the system size will determine whether or not we have decay from MQST.

In the following it is easier to work with the system size in terms of the coherence length ($\ell=L/\xi$). It is also convenient to measure the momentum in units of $1/\xi$, so that the allowed momentum modes are given by $q_n=2\pi n/\ell$ with an integer $n$~\endnote{The precise values of allowed momenta are determined by boundary conditions, which we assumed to be periodic for the purpose of this paper. Qualitatively the picture should not change if we use any other boundary conditions.}. The characteristic momentum mode that is resonant with the zero momentum mode oscillations is given by $q_c=1/\sqrt{2}$ in these units. The condition for MQST decay is simply
stated that for a given system size $\ell$, there exists at least one value of the integer $n$ such that
\begin{equation}
q_c\left(1-\frac{\sqrt{3}\lambda}{2}\right)<\frac{2\pi n}{\ell}<q_c\left(1+
\frac{\sqrt{3}\lambda}{2}\right). \label{eq:size}
\end{equation}
If this condition is not satisfied for any $n$ then MQST is stable even for a 1D system.

The requirement Eq.~\ref{eq:size} implies that there is a minimum system size $\ell_{\textrm{min}}$ below
which MQST is always stable:
\begin{equation}
\ell_{\textrm{min}} = \frac{2\pi}{q_c\left(1-\frac{\sqrt{3}\lambda}{2}\right)}. \label{eq:lmin}
\end{equation}
For small $\lambda$ to a very good approximation we find that this is equivalent to the minimum system size (in original units)
$L_{\textrm{min}} = 2\pi \sqrt{2} \xi$.
For all systems smaller than $\ell_{\textrm{min}}$ MQST is always stable and the effect of
quantum fluctuations can be ignored.  We then have a situation identical to the Josephson junction where MQST was originally studied, where only the dynamics in a single momentum mode is important in each wire.

We also note that there exists a system size $\ell_{\textrm{thresh}}$ above which MQST
is always unstable and the dynamical instability is always present. This corresponds to the situation whene there is always at least one solution satisying Eq.~(\ref{eq:size}). The condition that determines the threshold
length is that the spacing between adjacent momentum modes (that is $2\pi/\ell$) is smaller than the size of the dynamically unstable region in momentum space (see Figure~\ref{fig:imag}). The threshold length is given by
\begin{equation}
\ell_{\textrm{thresh}} = \frac{2\pi}{\sqrt{3}\lambda q_c},
\label{eq:lthresh}
\end{equation}
or in original units it is just
$L_{\textrm{thresh}} = \frac{2\pi}{\lambda}\sqrt{\frac{2}{3}}\xi$,  
so the threshold length is of the order of $\xi/\lambda$.  The smaller we make the tunneling
between the wires, the larger the threshold length will be.

Finally, there is a range of intermediate system sizes between the minimum size
$\ell_{\textrm{min}}$ below which MQST is always stable,
and the threshold size $\ell_{\textrm{thresh}}$ above which MQST is always unstable.
For system sizes $\ell_{\textrm{min}}<\ell<\ell_{\textrm{thresh}}$,
we find that by changing the system size we alternate between intervals where MQST
is stable and intervals where MQST is unstable. Moreover in unstable intervals there is always a single unstable mode. It is easy to see that unstable intervals in $\ell$ (for intermediate system sizes) are centered around
\begin{equation}
\ell_n = \frac{2\pi n}{q_c},
\label{eq:elln}
\end{equation}
where $n$ is an integer.  If we define $n_{\textrm{max}}$ as the largest value of $n$ for which
$\ell_n<\ell_{\textrm{thresh}}$ is satisfied, then we need to have
$n_{\textrm{max}} < \frac{1}{\sqrt{3}\lambda}$.
We also note that the size of the interval in $\ell$ where MQST is unstable increases linearly with $n$ so that the $n$-th unstable region corresponds to the following interval for $\ell$:
\begin{equation}
\frac{4\pi\sqrt{2}n}{2+\sqrt{3}\lambda} < \ell < \frac{4\pi\sqrt{2}n}{2-\sqrt{3}\lambda}.
\label{eq:unst}
\end{equation}
For $n>n_{\textrm{max}}$, the size of the unstable regions become so large that they overlap with each other and the system is always unstable.

Because in the intermediate regime $\ell_{\rm min}<\ell<\ell_{\rm thresh}$ there is at most one unstable mode, we should be able to observe a very sharp momentum distribution develop in unstable regions. Furthermore we can control how quickly the system decays from MQST; specifically we can set the system size such that the rate of decay has a maximum. This happens if the system parameters are such that the unstable mode perfectly matches the resonance condition. Clearly by decreasing the tunneling $\lambda$ we can make resonance conditions sharper and the intermediate region larger. To achieve the maximal growth rate in the population of the resonant mode (and hence the maximum number of particles transfered to this mode) we need to choose $\ell=\ell_n = 2\pi n/q_c$. Note that by picking such $\ell_n$ we are also guaranteed to get a momentum state that falls within the \emph{second} unstable region $q=2q_c$ in momentum space (see Figure~\ref{fig:imag}).  Recall, however, that the rate of growth here is of the order of $\lambda$ smaller than the corresponding one in the first unstable region, so we may neglect its effect if $\lambda$ is small
enough, which it is in cases where we initially have MQST. 
Let us finally mention that if the system is in the intermediate size regime, then according to Eq.~(\ref{eq:unst}) (see Figure~\ref{fig:imag} as well) one can easily tune the system from stable to unstable MQST by tuning $\lambda$ (or equivalently $J$).

The analysis we presented so far neglects the interactions between different modes. In particular, it neglects the damping of the zero momentum mode oscillations, which is inevitable due to creating high energy excitations in the system. This damping will destroy resonant population of the unstable modes and thus will lead to eventual saturation of the exponential growth. At longer times one can expect that the energy stored in the unstable momentum mode will be redistributed among other modes and the system will eventually thermalize (or at least reach some steady state). The two important questions are (i) What is the maximum achievable occupation of the resonant modes? and (ii) What is the time scale at which the sharp distribution disappears and the system thermalizes? Later when we do numerics, we will see that higher $n$ decreases the thermalization time for the system and decreases the maximum occupancy of the resonant modes. Therefore by picking smaller system sizes we can delay thermalization. If we go to the extreme case and pick the smallest system size, so that only the first excited momentum mode falls within the dynamically unstable region, we do not observe thermalization in our numerics for the time scales considered.  In particular, in this regime we do not see equilibration of particles between the two wires, and the system seems to remain self trapped for very long times. In this extreme case the sharp momentum distribution is stable, and the system
seems to reach a steady state in population of the nonzero momentum modes.  We do observe, however, a slow leak of atoms from the zero momentum mode of the full wire to the zero momentum mode of the empty wire, as seen in Figs.~\ref{fig:L1emp} and ~\ref{fig:L1full}.  Only the zero and first excited momentum modes are important in the dynamics. This is in contrast to larger system sizes (with modes in the unstable region), where the sharp momentum distribution disappears, the population difference between wires disappears, and the system thermalizes at much shorter times scales.

In general thermalization times decrease as the system size increases. If we pick system sizes
above $\ell_{\textrm{thresh}}$, MQST always
decays, and if $\ell$ is large enough we will always have multiple modes fall within the
dynamically unstable region.  If enough modes fall within
this region we will no longer have a sharp momentum distribution in intermediate times,
and in fact will not even be able to discern that there is MQST
in the system.  The time scale for MQST decay is on the order of $J$ if only one,
or at most a few modes fall within the unstable region.  This time
scale decreases as we make $\ell$ larger until eventually it becomes on the order of
$\mu$, the frequency of zero momentum mode oscillations.  At this point, we
can no longer claim that we had MQST in the first place.  So in the thermodynamic
limit there is never any MQST in 1D systems.

 Let us note that while our analysis directly applies to uniform systems, one can expect that qualitatively the results will be the same in a parabolic trap. In this case the noncondensate eigenmodes will not be plane waves but wavepackets oscillating with characteristic momentum, which will be quantized. If the resonant conditions are satisfied we expect MQST decay while if they are not, MQST will be stable.

\section{Numerics  \label{sec:numerics}}
In this section we will present the numerical results.  Our system is described by the
microscopic Hamiltonian (\ref{eq:nuh}).  We will work with weakly interacting systems. In this regime the initial condensate fraction is large and the resonant mode population is most pronounced. Also for weak interaction we can reliably use truncated Wigner approximation (TWA) method to simulate the dynamics (see Refs.~\cite{walls-milburn, gardiner-zoller, steel, polkovnikov2003, blakie_08, polkovnikov2009} for details of this method). TWA gives the leading order in expansion of quantum dynamics in \emph{quantum fluctuations} around the classical (Gross-Pitaevskii) limit. The classical limit can be strongly interacting in the sense that classical dynamics is intrinsically nonlinear. It can be shown that in TWA the classical equations of motion do not change but the initial conditions for classical fields become fluctuating and distributed according to the Wigner function. These fluctuations are crucial for our problem since without them non-zero momentum modes are never excited. We emphasize that we use the term ``classical'' for Gross-Pitaevskii equations rather than mean-field commonly used in literature. It is important to realize that in TWA each classical trajectory does not describe any mean-field, it rather describes characteristics along which the Wigner function is approximately conserved. Only in the classical limit where there are no fluctuations and all particles are described by a single condensate mode Gross-Pitaevskii equations describe mean-field dynamics. Effectively the Wigner function in TWA gives the quasiprobability distribution for classical fields determined by the initial state of the system, pure or mixed. At higher orders, there will be also corrections to the classical trajectories, but in the regime of large $K$ we are interested in, those are unimportant~\cite{polkovnikov2003, polkovnikov2009}. 

The implementation of TWA is straightforward. The dynamics of the system is described by Gross-Pitaevskii equations of motion for the following Hamiltonian
\begin{eqnarray}
H_{\rm cl} &=& \sum_{i=1}^2 \int_0^L dx \, \left(\frac{1}{2m} \partial_x \Psi_i^{\star}\partial_x \Psi_i
+\frac{g}{2} \Psi_i^{\star}\Psi_i^{\star}\Psi_i\Psi_i \right)
\nonumber\\
&& -J\int_0^L dx \, \left(\Psi_1^{\star}\Psi_2+\Psi_2^{\star}\Psi_1\right), \label{eq:nuhc}
\end{eqnarray}
where we simply replaced the quantum operators in Eq.~(\ref{eq:nuh}) with classical fields.
We note that in principle one has to use the Weyl symbol of the Hamiltonian in Eq.~(\ref{eq:nuhc}). However, for spatially uniform quartic interactions the difference between (\ref{eq:nuhc}) and the Weyl symbol is given by a constant term proportional to the total number of particles. In equations of motion it yields an overall phase, which is not important. The Gross-Pitaevskii equations corresponding to this Hamiltonian are obtained in the standard way:
\begin{equation}
i\partial_t\Psi_i(x)={\delta H_{\rm cl}\over\delta\Psi^\star(x)}.
\label{eq:gp}
\end{equation}
Note that the same equation can be obtained by introducing an analogue of Poisson brackets for the coherent states and writing standard classical Hamilton equations of motion~\cite{polkovnikov2009}.

To implement TWA we need to solve equations (\ref{eq:gp}) supplemented by random initial conditions distributed according to the Wigner transform of the initial density matrix. Since we are dealing with an interacting system in the initial state, finding the Wigner transform is not trivial by itself. One possibility to overcome this difficulty is to find the initial state within the Bogoliubov's approximation and then find the Wigner transform of the corresponding wave function~\cite{blakie_08}. We will employ a different route. Namely we start with a noninteracting full wire ($g=0$) where it is straightforward to find the Wigner transform of the ground state. Then we slowly (adiabatically) turn on interactions in the system (keeping the wires decoupled) until we reach full interaction strength. Since quantum fluctuations remain weak in the whole range of $g$ this procedure gives us the correct initial state for the interacting system (we checked that this procedure is very accurate on a small lattice system comparing such adiabatic ramp with exact simulations). We note that in this setup it is actually very hard to achieve adiabatic limit in large 1D systems~\cite{adiabnp}. However, this does not represent a real issue for us since we confine ourselves to mesoscopic system sizes. Once the desired interaction strength is reached we suddenly turn on coupling between the wires and analyse the subsequent dynamics.

\begin{figure}[pt]
	\centering
		\includegraphics[scale=0.7,angle=270]{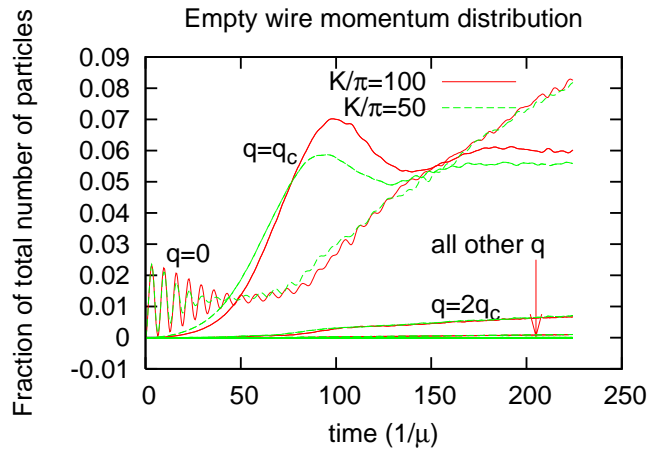}
		\caption{(Color online) Momentum distribution in the empty wire for $L=2\pi/q_c$, the smallest system size for which MQST decay can be observed, and $\lambda=0.075$.  The two sets of curves correspond to the Luttinger parameters $K/\pi=50$ and $K/\pi=100$ respectively. Short time small amplitude oscillations of $q=0$ modes correspond to MQST. At intermediate times resonant modes at $q=q_c$ are populated and the momentum distribution is sharply peaked at $q=\pm q_c$ reaching the maximum population of about $8\%$ of particles per mode. The maximum is bigger for weaker interactions (larger $K$). After the saturation of the resonant mode we observe very slow leak of atoms into the empty wire in the zero momentum mode with a large population imbalance persisting for very long times. Note that only few modes ($q=0,\pm q_c,\pm 2 q_c$) are
involved in the dynamics. 
\label{fig:L1emp} }
\end{figure}

\begin{figure}[pt]
	\centering
		\includegraphics[scale=0.7,angle=270]{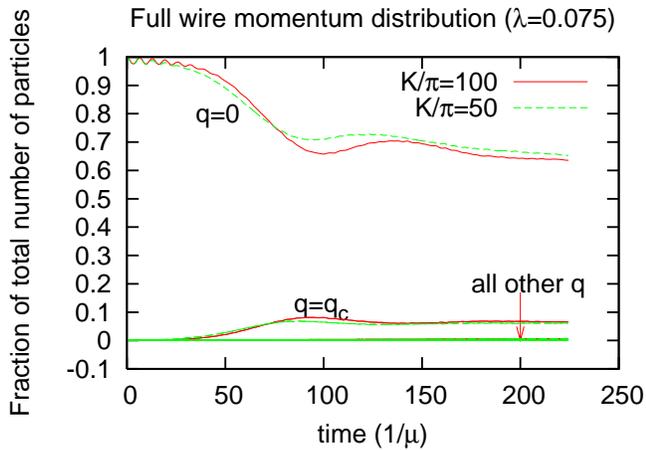}
		\caption{(Color online) Same as Figure~\ref{fig:L1emp} but for the full wire. Again, only the lowest modes are involved in the dynamics ($q=0,\pm q_c,\pm 2 q_c$) and we find no equilibration in these time scales.  \label{fig:L1full} }
\end{figure}

We consider weakly interacting systems with $K/\pi= 50, \,100$ so that we are certainly in the semiclassical limit. Since $K/\pi=\rho_0\xi$ we see that this corresponds to the regime of roughly $100$ particles within the healing length. The minimal length of the wire required to see decay of MQST is $L_{\rm min}=2\pi\sqrt{2}\xi\approx 9\xi$ (see Eq.~(\ref{eq:lmin})) thus we see that this parameter regime requires of the order of few hundred particles in the full wire. The effect will persist for smaller values of $K/\pi\gtrsim 10$ so it can be observed with smaller number of particles per wire, though for smaller $K$ the momentum distribution becomes not so sharp. We will choose small couplings $\lambda=0.075, \, 0.05$ so that the two relevant time scales in the system ($1/J$ and $1/\mu$) are well separated and we can clearly distinguish the MQST part of the dynamics at short times from the sharp momentum distribution that develops at times $\sim 1/J$.  Finally, we will work mostly with system sizes $\ell\equiv L/\xi=\ell_1, \, \ell_{10}<\ell_{\rm thresh}$ (see Eqs.~(\ref{eq:lthresh}), (\ref{eq:elln})), so that we are in the regime with exactly one momentum mode falling within the dynamically unstable region. We also choose system sizes that maximize the imaginary part of the frequency of the unstable mode.  This gives us the largest transfer of particles to these modes.

Let us first discuss in more detail what happens for $L=L_1=2\pi/q_c$.  We have plotted the
results for momentum mode occupation for all momentum modes for times up to $t=225/\mu$ in Figure~\ref{fig:L1emp} for the empty wire, and Figure~\ref{fig:L1full} for the full wire.  We choose $\lambda=0.075$ and two different Luttinger liquid parameters $K/\pi=50$ and $100$. At short times we clearly observe  MQST, which is characterized by the high frequency, small amplitude oscillations in the zero momentum mode. The behavior
at short times is similar for both $K/\pi=50$ and $K/\pi=100$, but note that the
damping of oscillations is stronger for smaller $K$, which is not surprising since
this corresponds to stronger quantum fluctuations. Then at times $\sim 1/J$, we find a large
transfer of particles to the characteristic momentum modes $q_c$, which for the given system size is just the first excited momentum mode in the system. A larger fraction of particles is transferred for larger $K$,
which is expected since larger interactions preclude a large population from developing
at any mode. We also clearly observe transfer of particles to the $2q_c$ mode.  Recall from our previous discussion that there is also an unstable region around $2q_c$, but the rate of transfer
there is of order $1/\lambda$ smaller.  All other momentum modes remain virtually
unpopulated at all times shown, so the dynamics is confined only to these
first few modes, and we find that this greatly delays the onset of thermalization
(see Figure~\ref{fig:L1full}). Note that at later times the occupation
in $q_c$ saturates at about $8\%$ of the total number of particles, while the population in the zero momentum mode monotonically increases and eventually surpasses the population in the characteristic modes.  The system is
thermalizing, albeit very slowly.

\begin{figure}[pt]
	\centering
		\includegraphics[scale=0.7,angle=270]{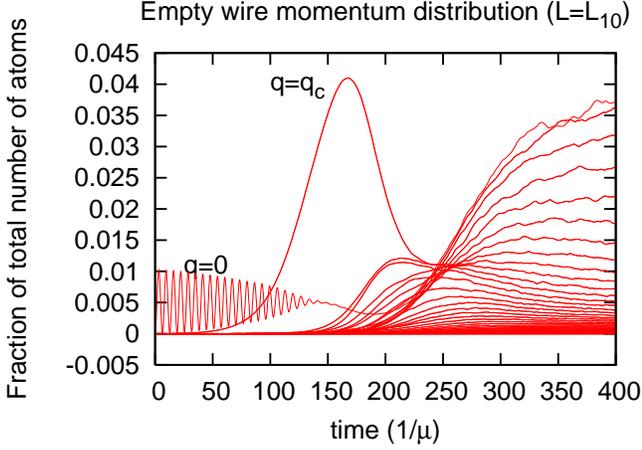}
		\caption{(Color online) Momentum distribution in the empty wire where we have $L=20\pi/q_c$,
$\lambda=0.05$, and $K/\pi=100$. At short times there is clear MQST then at times $\sim 1/J$ we find a very sharp momentum distribution as particles are transferred to the characteristic momentum modes.  The system size is picked such that the characteristic mode corresponds to the tenth excited momentum mode in the system.  At later times, particles are distributed in many momentum modes and the system thermalizes. This thermalization is accompanied by MQST decay.
\label{fig:L10emp} }
\end{figure}

\begin{figure}[pt]
	\centering
		\includegraphics[scale=0.7,angle=270]{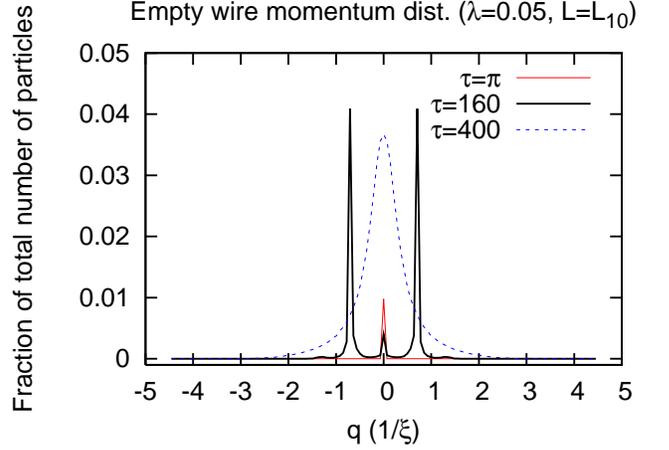}
		\caption{(Color online) Momentum distribution in the empty wire at several different times.
We use the same parameters as in Figure~\ref{fig:L10emp}. At time $\tau=\pi$ (corresponding to half the period of MQST dynamics) only the zero momentum mode is populated.  At time $\tau=160$ we find a very sharp momentum distribution
centered around characteristic modes $\pm q_c$.  This time corresponds to the peak population
in $\pm q_c$ modes. Note that at this time the population in $\pm q_c$ by far eclipses the population of the zero momentum mode. Finally at time $\tau=400$ we observe a very broad thermal-like momentum distribution centered around $q=0$. 
\label{fig:L10mom} }
\end{figure}

\begin{figure}[pt]
	\centering
		\includegraphics[scale=0.7,angle=270]{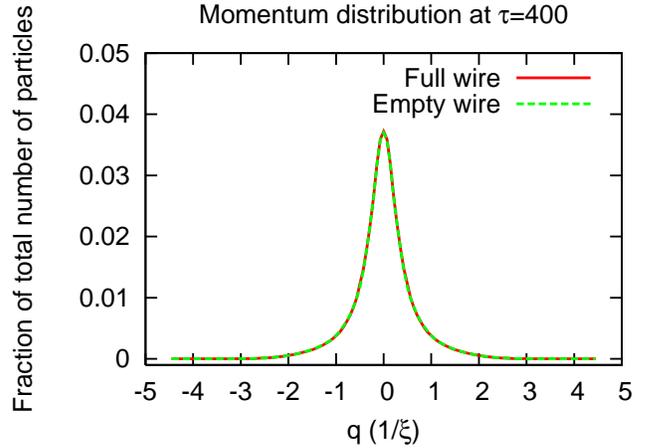}
		\caption{(Color online) Plot of the momentum distributions in both wires at time $\tau=400$.
Parameters are the same as in Figure~\ref{fig:L10emp}. Note that the two curves are practically indistinguishable.  The system has thermalized at this point. \label{fig:L10fin} }
\end{figure}

\begin{figure*}[floatfix]
	\centering
                \includegraphics[scale=0.7,angle=270]{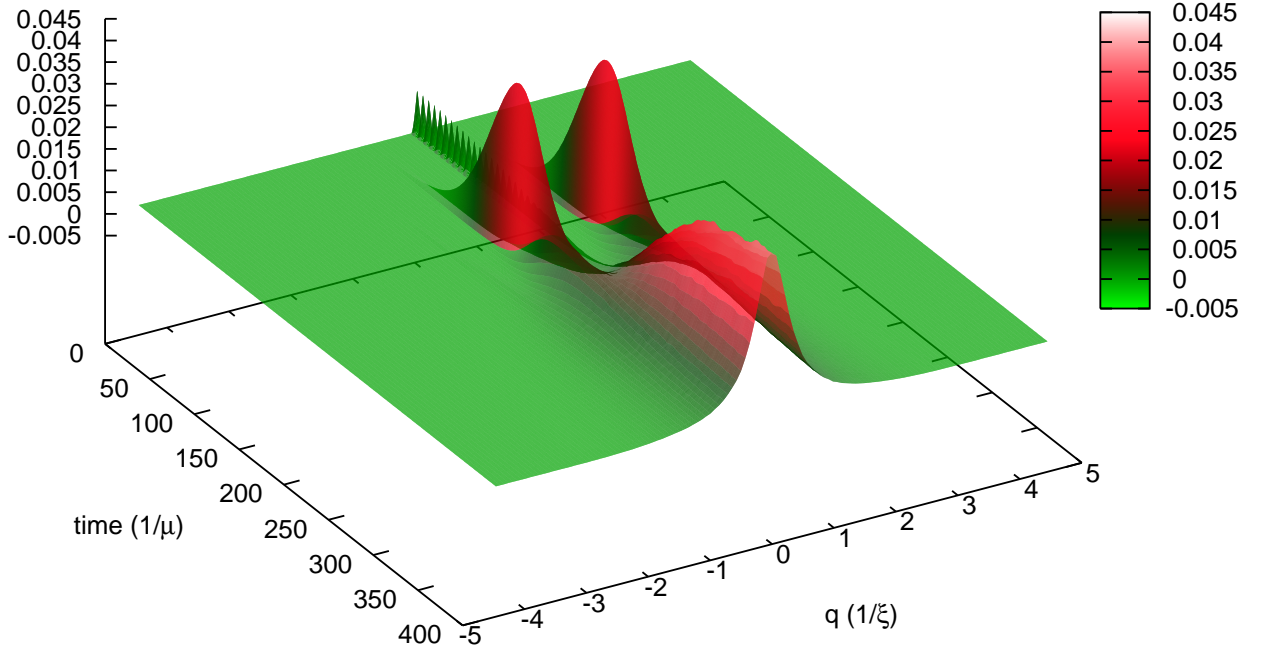}
		\caption{(Color online) A 3D plot of the momentum distribution in the empty wire vs
time and momentum. The parameters are shown in the plot and are the same as in Figure~\ref{fig:L10emp}.
One can clearly distinguish three stages in the dynamics corresponding to MQST, unstable resonant dynamics producing a sharply peaked momentum distribution, and finally thermalization.
  \label{fig:nice} }
\end{figure*}

Let us now focus on a larger system size $L=L_{10}=20\pi/q_c$.  This system size is such that
the tenth excited momentum mode falls within the unstable region. We pick $K/\pi=100$ and $\lambda=0.05$.  This choice of parameters allows us to delay thermalization so that we can clearly distinguish what happens in different time scales in the system. The results for the empty wire momentum occupation are shown in Figure~\ref{fig:L10emp},
(a 3D plot is shown in Figure~\ref{fig:nice}). At short times we once again observe the zero momentum mode oscillations present in MQST.  We observe the transfer of particles to the characteristic mode $q_c$ at intermediate time scales $\sim 1/J$.  Note that when the population at $q_c$ peaks, the momentum distribution is very sharply peaked around $\pm q_c$ as we see in Figure~\ref{fig:L10mom}. The system thermalizes at later times. In Figure~\ref{fig:L10mom} we compare the momentum distribution of the empty wire at different times.
At $t=\pi/\mu$, which corresponds to a half period in MQST oscillations, we find a small number of particles at zero momentum while all other modes are virtually unoccupied.  At intermediate times ($t=160/\mu$, which is of
the order of $1/J$) we find the momentum distribution very sharply peaked around $\pm q_c$ (with a small peak at $q=0$), while all other modes remain unoccupied.  At late times ($t=400/\mu$) the system
shows a thermal momentum distribution centered around $q=0$. Note that the thermalization time scale is much smaller here than when $L=L_1$.  Also the height of the peak when the population at $q_c$ is maximum
is higher than the maximum height of the thermalized momentum distribution.  This fact
taken with the fact the the distribution is very sharp when the population at $q_c$ peaks should make the intermediate sharp momentum distribution state easily distinguishable from the final thermalized state.
For the latest times shown $t=400/\mu$, we compare the momentum distribution of both
wires in Figure~\ref{fig:L10fin}. Note that they are virtually indistinguishable at this point so the system fully thermalized.

There is an important difference between dynamics in short systems with $\ell=\ell_1$ and longer systems with $\ell=\ell_n$, where $2<n<n_{\rm thresh}$ (compare Figs.~\ref{fig:L1emp} and \ref{fig:L10emp}). While in both cases there is only one unstable mode which gets populated at intermediate times $\sim 1/J$ the longer time dynamics is clearly different. In both cases the population of the resonant mode $q=q_c$ saturates at a value of the order of a few percent of the total population. At later times in the shorter wire the population remains confined essentially to the condensate (zero momentum) and $q=\pm q_c$ momentum modes, while in longer wires other momentum modes get excited and the system thermalizes. In order to observe sharp momentum distribution in this case it is necessary that the thermalization time is longer than $1/J$, the time scale at which instability develops. We find that in order to ensure this is the case the following condition has to be satisfied:
\begin{equation}
1-\left(\frac{\xi}{L_n}\right)^{1/2K} \lesssim \frac{4J^2}{\mu^2}=4\lambda^2. \label{eq:nj}
\end{equation}
The right side of Eq.~\ref{eq:nj} is just the amplitude of oscillations of the $q=0$ mode particle tunneling
at short times (while MQST is present), and the left hand side gives us roughly the depletion of particles from the zero momentum mode in the full wire. The basic idea behind Eq.~(\ref{eq:nj}) is that the amplitude of MQST zero momentum mode oscillations should be easily distinguishable from the depletion background due to interactions in order to have well defined MQST dynamics at short times. We verified numerically that if the inequality in Eq.~(\ref{eq:nj}) is not satisfied, the momentum distribution is broader and the system thermalizes before the peaks can fully develop, making them harder to observe. Thus for $\lambda=0.075$, the RHS of Eq.~(\ref{eq:nj}) is $0.025$ so we can have
a few percent depletion at most in the full wire. If we increase $\lambda$ then we can tolerate larger depletion. However, $\lambda$ can not be too large otherwise oscillations of the condensate mode are unharmonic and the instability is suppressed. In practice we find that for $\lambda > 0.1$ the system does not produce as sharp a momentum distribution. Even though the peaks are still distinguishable, their amplitude is smaller and they acquire some width compared to the case of smaller $\lambda$ satisfying Eq.~(\ref{eq:nj}).

\begin{figure}[pt]
	\centering
		\includegraphics[scale=0.65,angle=270]{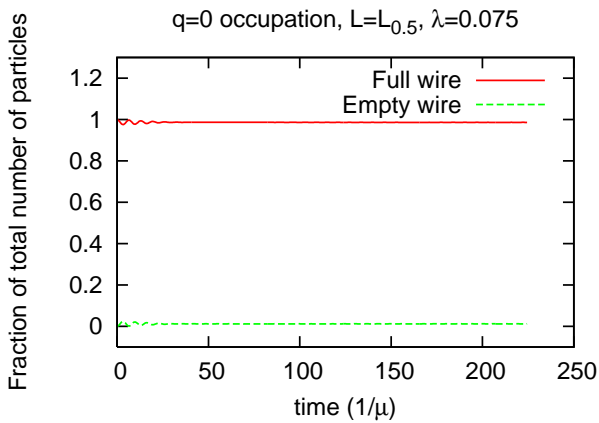}
		\caption{(Color online) Plot of zero momentum occupation for both full and empty wires for
$L=\pi/q_c$.  This is half the smallest system size necessary to have MQST decay.  Note that the system remains self trapped, but the MQST oscillations in the zero momentum modes are damped, unlike the usual case of MQST in a Bosonic Josephson junction.
 \label{fig:L0p5} }
\end{figure}

\begin{figure}[pt]
	\centering
		\includegraphics[scale=0.65,angle=270]{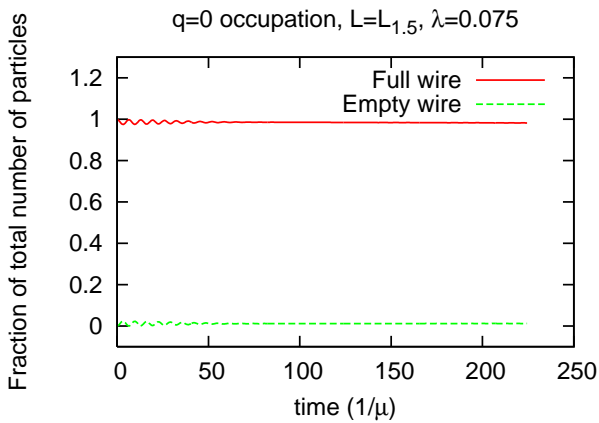}
		\caption{(Color online) Plot of zero momentum occupation for both full and empty wires for
$L=3\pi/q_c$.  According to our
previous analysis this system should exhibit no MQST decay.  This system size falls in an
interval of MQST stability
that is sandwiched between system size intervals where MQST is unstable, which shows
it is possible to tune the system
size to alternate between regions of MQST stability and MQST decay.
Note that the system remains self trapped, but the MQST oscillations in the
zero momentum modes are damped, unlike the usual case of MQST in a Bosonic Josephson junction.
 \label{fig:L1p5} }
\end{figure}

For completeness we show the momentum distribution for system sizes smaller than the
minimum required for MQST decay in Figure~\ref{fig:L0p5} (specifically $L=\pi/q_c$, which
is half the minimum size required for MQST decay) and show that MQST is indeed stable in such a system.  Note that even though the system remains self trapped, the zero momentum mode oscillations are damped, which does not happen in a
single Josephson junction.  The source of damping likely originates from coupling to non-resonant modes, which are always present in 1D (note that this system size still exceeds interparticle distance by two orders of magnitude). We also show that MQST is stable for $L=3\pi/q_c$ in Figure~\ref{fig:L1p5}. This system size is exactly in the middle between the first and second instabilities. This shows that it is possible to tune the system
size such that we alternate between MQST stability and MQST decay.  Note that there is no observable thermalization in the system for very long times. Finally let us mention that for $L\gg L_{\rm thresh}$ (see Eq.~(\ref{eq:lthresh})) where there are many unstable modes, the system thermalizes so quickly that there is no MQST dynamics present at all, and a sharp momentum distribution never develops.

\begin{figure}[pt]
	\centering
		\includegraphics[scale=0.65,angle=270]{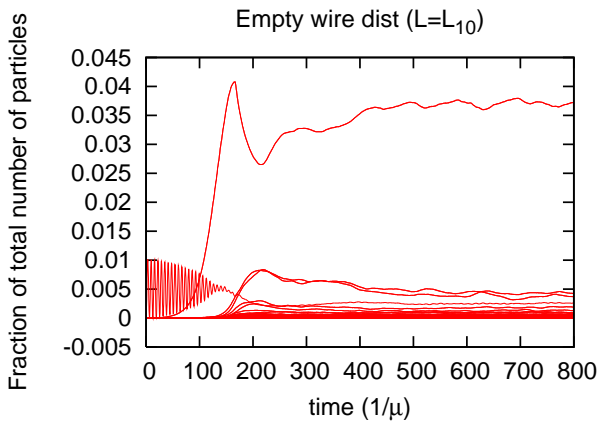}
		\caption{(Color online) Momentum distribution in the empty wire where we have $L=20\pi/q_c$,
$\lambda=0.05$, and $K/\pi=100$. These
are the same parameters used in Figure~\ref{fig:L10emp}, except now we decouple the wires at
$t=140/\mu$, which is roughly when the
population in the characteristic modes peaks. Note that the momentum distribution remains
highly peaked around the characteristic momentum modes.
Since the system is now integrable, we do not observe thermalization.
\label{fig:L10dec} }
\end{figure}

Consider now the very interesting possibility of decoupling
the wires immediately after obtaining the sharp momentum distribution in the empty wire.  Consider once again the parameters used in Figure~\ref{fig:L10emp}, where we have $L=20\pi/q_c$, $K/\pi=100$, and $\lambda=0.05$.  Note that the population in the characteristic momentum modes peak around $t\simeq 140/\mu$. To preserve this highly nonequilibrium momentum distribution at this moment we can suddenly decouple the wires. The resulting dynamics is shown in Figure~\ref{fig:L10dec}.  Clearly the momentum distribution remains peaked around the characteristic modes $q_c$, which is very different from the dynamics shown in Figure~\ref{fig:L10emp} where the wires remain coupled.  The system does not equilibrate in the usual sense. This is not surprising since after decoupling the wires we are left with a 1D interacting Bose gas of the Lieb-Liniger type (see Ref~\cite{lieb1963}),
which is an integrable system solvable by Bethe ansatz.  The integrability of the system precludes thermalization and as a consequence we observe that the sharp momentum distribution is robust.  Through the procedure considered here, we have found a way to construct a 1D system with a stable sharp momentum distribution around nonzero characteristic momentum modes, which may be used itself in interesting applications. Note that in the \emph{empty} wire almost all atoms occupy $\pm q_c$ modes, which is a highly nonequilibrium state akin to having \emph{large} counterpropagating currents in the same wire.  In a usual nonintegrable system one expects that collisions between counterpropagating particles would thermalize the system and the large currents would dissipate, but integrability in the system considered here protects these currents and we observe no current decay (see also Ref.~\cite{kinoshita}).

Before we conclude this section, let us make a few comments on the regime of applicability of our results. We focused mainly in the case of fairly large $K/\pi=50, \, 100$ since in these cases one has the very interesting situation where we have a \emph{very large} transfer of atoms to the characteristic modes and therefore a \emph{very sharp} momentum distribution develops before thermalization sets in, which makes this choice of parameters very interesting from a theory point of view.  However, qualitatively our conclusions remain robust to much stronger interactions as long as  Eq.~(\ref{eq:dep}) is satisfied. For smaller $K$ we observe smaller transfer of particles to the characteristic modes and thus not so sharp momentum distribution. We find that for $K/\pi \geq 10$ the intermediate time momentum distribution is easily discernible from the final thermal distribution. For comparison, let us mention that for $K/\pi=20$ and $\ell=20\pi/q_c$ (so the characteristic mode is the tenth excited momentum mode), we get about $1.6\%$ of atoms transferred to \emph{each} characteristic mode (i.e. $\sim 3\%$ for both $\pm q_c$), compared to $4\%$ for each characteristic mode for $K/\pi=100$, and the distribution remains fairly sharp for this case. For $K/\pi=10$, however, the percentage transferred to each characteristic mode goes down to $0.6\%$ and the distribution becomes noticeably broader.

At the same time the nonmonotonic dependence of MQST decay on system size remains \emph{very} robust as long as $K/\pi\geq 10$. Thus for $K/\pi \gtrsim 20$, the system size dependence is practically indistinguishable from what happens at larger $K$ (i.e. $K/\pi=100$, see Figures~\ref{fig:L0p5} and \ref{fig:L1p5}). In particular, we still see essentially \emph{no} decay from MQST if the system size is such that there are no unstable modes present, and if there are unstable modes present in the system we see decay of MQST in a similar fashion to what happens for larger $K$. For $K/\pi=10$, we still observe nonmonotonic dependence of MQST decay with system size, however unlike the situation with larger $K$ (see Figs.~\ref{fig:L0p5}, \ref{fig:L1p5}) there is some small yet noticeable decay of MQST even when the resonance condition is not satisfied.  For $K/\pi\simeq 5$, we find that the nonmonotonic behavior of MQST decay with system size is no longer present.

\section{Discussion \label{sec:summary}}
In this paper we studied the breakdown of MQST for two coupled 1D interacting Bose gases
with a large population imbalance. There are certain constraints that the system must
meet in order for it to initially exhibit self trapping.  In particular, the system size should be mesoscopic such that the initial depletion from the quasicondensate modes remains relatively small. Also the interactions in the system should be weak corresponding to large Luttinger liquid parameter $K\gtrsim 30$. If these conditions are satisfied the initial self-trapping occurs if the ratio of the tunneling coupling $J$ to the chemical potential in the full wire $\mu$ is sufficiently small: $\lambda=J/\mu<1/4$.

Under these conditions the system exhibits MQST at short times.  MQST is characterized by small amplitude (order $\lambda^2$) tunneling of zero momentum mode (condensate) particles between two 1D wires. The frequency of these characteristic MQST zero momentum mode oscillations is $\mu$, the chemical potential difference between wires.  At intermediate times (of order $1/J$, where $J$ is the tunneling energy) we find that MQST can decay.  The mechanism of this decay is the resonant coupling between the oscillating zero momentum mode fields and creation of quasiparticle pairs of characteristic momenta $\pm q_c$ with the energy matching the oscillation frequency. These excited modes have a very high momentum on the order of the inverse healing length in the full wire (or equivalently with the energy of the order of $\mu$).  Through this mechanism, the system can develop a very sharp momentum distribution around $\pm q_c$ at intermediate times. At later times, the system generally reaches a steady state at which point the momentum distribution in the two wires is identical.

Since we are working with mesoscopic systems, we find that the system size puts a very strong constraint on whether or not we observe MQST decay, due to the fact that we have a discrete set of allowed momentum modes.  Simply put, if there are no modes that satisfy the resonance condition (within a tolerance of order $J$), then
MQST is stable even in 1D systems and the population imbalance survives for very long times. A corollary to this is that there is a minimum system size $L_{\textrm{min}} \simeq 2\pi/q_c$ below which MQST is always stable, and a system size $L_{\textrm{thresh}} \simeq 2\pi /\sqrt{3}\lambda q_c$ above which one always has MQST decay.
For intermediate system sizes between these two, in order to have MQST decay we must confine ourselves to system sizes falling in a region centered around discrete values $L_n \simeq 2\pi n/q_c$, with these regions having a size
of order $\lambda n$.  By picking exactly $L_n = 2\pi n/q_c$ (with $L<L_{\textrm{thresh}}$), so that the characteristic
momentum mode is exactly resonant with the tunneling rate of zero momentum mode atoms,
one can maximize the rate of MQST decay and maximize the transfer
of particles to the characteristic modes $\pm q_c$ before thermalization sets in.
For the very special case where one picks the smallest system size where MQST decay is possible ($L_1 = 2\pi /q_c$) we find that the system reaches a steady state only at very long times, since only very few system modes participate in the dynamics.

Interestingly, by taking advantage of the MQST decay mechanism, the system considered in this paper may be used to construct
stable states with a sharp momentum distribution around nonzero characteristic momentum modes.
One simply decouples the two wires after enough particles are transfered to the empty wire
before thermalization sets in, i.e. while the momentum distribution is sharp. The decoupled
wires are integrable systems (see Ref.~\cite{lieb1963}), and the system does not thermalize after
decoupling.  The empty wire in particular has almost all particles occupying the
characteristic modes $\pm q_c$. This is akin to having large counterpropagating currents in
the same wire, where interestingly we observe no current decay due to the integrability
of the system.

We briefly mention here that there exists another way of preserving the sharp momentum distribution. Instead of decoupling the wires (while the momentum distribution is still sharply peaked around characteristic momenta $\pm q_c$), one can also achieve the same effect by introducing a \emph{large} uniform potential bias $V$ (with $V \sim \mu$) between the two wires, the empty wire being at a \emph{lower} potential (if instead the empty is at a higher potential, then the self trapping trivially survives and there is not much transfer between wires since it is more favorable for them to stay in the full wire).  The behavior at short times is essentially identical to the nonbiased case considered in this paper, the only difference being the trivial rescaling of the parameters in the biased case.  There is however a stark difference in behavior between the two cases at times greater than $O(1/J)$. In this situation, the sharp momentum distribution in the empty wire remains robust, and MQST survives for long times.  The height of the peaks in the momentum distribution at first sight seems to saturate at the same value as before (i.e. $4.5$ percent of total $N$ for $K/\pi=100$ and $L=L_{10}$). In our numerics, however, we do notice a slow leak of particles from the zero momentum mode in the full wire to \emph{nonzero} momentum modes in the empty wire surrounding, and including, the highly populated characteristic modes $\pm q_c$. This effect slowly increases the number of particles in the empty wire, while mostly maintaining the sharpness of the momentum distribution. There is noticeable broadening developing at later times, and we expect that the peaks will once again saturate to a value similar to that found in the nonbiased case before thermalization sets in, but this requires further investigation. Interestingly we observe essentially no transfer to zero momentum modes in the empty wire, in stark contrast to what happens with no bias. Because of this slow decay that persists, we expect that at time scales much larger than $1/J$ (beyond what we have investigated numerically) MQST eventually is destroyed, but the time scales for this seem to be very high. More work needs to be done in the biased case. Note once again that in this scenario the wires remain coupled at \emph{all} times $t>0$, the only thing we do is introduce an additional potential bias before coupling the wires at $t=0$. 

As a final remark we point out that in the setup we analyzed dynamics is particularly interesting for very weak but nonzero interactions (see Fig.~\ref{fig:nice}). In this case one can clearly see three stages of evolution: (i) macroscopic self trapping, (ii) resonant excitation of characteristic modes, and (iii) thermalization between the wires. As interactions get stronger the second stage gradually disappears and the sharp momentum distribution at intermediate times becomes less pronounced. Nevertheless, according to our numerical results, all three features can be observed for $K/\pi \gtrsim 10$, which is realistic for present experiments. Another important prediction of our analysis, which can be easily verified in experiments, is very strong and nonmonotonic dependence of the MQST decay time on the system size. In particular, we find that in the interval $L_{\rm min}<L<L_{\rm thresh}$ this decay time is relatively short $\tau\sim 1/J$ if the resonance condition is satisfied (i.e. there is an unstable mode) and very long if it is not satisfied. This result is valid even in the systems with larger interactions, where a narrow momentum distribution at intermediate times never forms.

\begin{acknowledgments}
We acknowledge helpful discussions with R. Barankov and A.~M.~Rey. We also thank I. Bloch for suggesting this problem and sharing some details on the experimental system. R.H. acknowledges helpful discussions
with D. W. Hutchinson on the implementation of TWA in continuum systems and acknowledges E. Dalla Torre for helpful comments on the manuscript. This work was supported by AFOSR YIP and Sloan Foundation.
\end{acknowledgments}

\bibliography{breakdown5.bib}

\begin{thebibliography}{32}
\expandafter\ifx\csname natexlab\endcsname\relax\def\natexlab#1{#1}\fi
\expandafter\ifx\csname bibnamefont\endcsname\relax
  \def\bibnamefont#1{#1}\fi
\expandafter\ifx\csname bibfnamefont\endcsname\relax
  \def\bibfnamefont#1{#1}\fi
\expandafter\ifx\csname citenamefont\endcsname\relax
  \def\citenamefont#1{#1}\fi
\expandafter\ifx\csname url\endcsname\relax
  \def\url#1{\texttt{#1}}\fi
\expandafter\ifx\csname urlprefix\endcsname\relax\def\urlprefix{URL }\fi
\providecommand{\bibinfo}[2]{#2}
\providecommand{\eprint}[2][]{\url{#2}}

\bibitem[{\citenamefont{Giovanazzi et~al.}(2000)\citenamefont{Giovanazzi,
  Smerzi, and Fantoni}}]{giovanazzi2000}
\bibinfo{author}{\bibfnamefont{S.}~\bibnamefont{Giovanazzi}},
  \bibinfo{author}{\bibfnamefont{A.}~\bibnamefont{Smerzi}}, \bibnamefont{and}
  \bibinfo{author}{\bibfnamefont{S.}~\bibnamefont{Fantoni}},
  \bibinfo{journal}{Phys. Rev. Lett.} \textbf{\bibinfo{volume}{84}},
  \bibinfo{pages}{4521} (\bibinfo{year}{2000}).

\bibitem[{\citenamefont{Levy et~al.}(2007)\citenamefont{Levy, Lahoud, Shomroni,
  and Steinhauer}}]{levy2007}
\bibinfo{author}{\bibfnamefont{S.}~\bibnamefont{Levy}},
  \bibinfo{author}{\bibfnamefont{E.}~\bibnamefont{Lahoud}},
  \bibinfo{author}{\bibfnamefont{I.}~\bibnamefont{Shomroni}}, \bibnamefont{and}
  \bibinfo{author}{\bibfnamefont{J.}~\bibnamefont{Steinhauer}},
  \bibinfo{journal}{Nature} \textbf{\bibinfo{volume}{449}},
  \bibinfo{pages}{579} (\bibinfo{year}{2007}).

\bibitem[{\citenamefont{Zapata et~al.}(1998)\citenamefont{Zapata, Sols, and
  Leggett}}]{zapata1998}
\bibinfo{author}{\bibfnamefont{I.}~\bibnamefont{Zapata}},
  \bibinfo{author}{\bibfnamefont{F.}~\bibnamefont{Sols}}, \bibnamefont{and}
  \bibinfo{author}{\bibfnamefont{A.~J.} \bibnamefont{Leggett}},
  \bibinfo{journal}{Phys. Rev. A} \textbf{\bibinfo{volume}{57}},
  \bibinfo{pages}{R28} (\bibinfo{year}{1998}).

\bibitem[{\citenamefont{Raghavan et~al.}(1999)\citenamefont{Raghavan, Smerzi,
  Fantoni, and Shenoy}}]{raghavan1999}
\bibinfo{author}{\bibfnamefont{S.}~\bibnamefont{Raghavan}},
  \bibinfo{author}{\bibfnamefont{A.}~\bibnamefont{Smerzi}},
  \bibinfo{author}{\bibfnamefont{S.}~\bibnamefont{Fantoni}}, \bibnamefont{and}
  \bibinfo{author}{\bibfnamefont{S.~R.} \bibnamefont{Shenoy}},
  \bibinfo{journal}{Phys. Rev. A} \textbf{\bibinfo{volume}{59}},
  \bibinfo{pages}{620} (\bibinfo{year}{1999}).

\bibitem[{\citenamefont{Albiez et~al.}(2005)\citenamefont{Albiez, Gati,
  F{\"o}lling, Hunsmann, Cristiani, and Oberthaler}}]{albiez2005}
\bibinfo{author}{\bibfnamefont{M.}~\bibnamefont{Albiez}},
  \bibinfo{author}{\bibfnamefont{R.}~\bibnamefont{Gati}},
  \bibinfo{author}{\bibfnamefont{J.}~\bibnamefont{F{\"o}lling}},
  \bibinfo{author}{\bibfnamefont{S.}~\bibnamefont{Hunsmann}},
  \bibinfo{author}{\bibfnamefont{M.}~\bibnamefont{Cristiani}},
  \bibnamefont{and} \bibinfo{author}{\bibfnamefont{M.~K.}
  \bibnamefont{Oberthaler}}, \bibinfo{journal}{Phys. Rev. Lett.}
  \textbf{\bibinfo{volume}{95}}, \bibinfo{pages}{010402}
  (\bibinfo{year}{2005}).

\bibitem[{\citenamefont{F\"olling et~al.}(2007)\citenamefont{F\"olling,
  Trotzky, Cheinet, Feld, Saers, Widera, M\"uller, and Bloch}}]{trotzky_07}
\bibinfo{author}{\bibfnamefont{S.}~\bibnamefont{F\"olling}},
  \bibinfo{author}{\bibfnamefont{S.}~\bibnamefont{Trotzky}},
  \bibinfo{author}{\bibfnamefont{P.}~\bibnamefont{Cheinet}},
  \bibinfo{author}{\bibfnamefont{M.}~\bibnamefont{Feld}},
  \bibinfo{author}{\bibfnamefont{R.}~\bibnamefont{Saers}},
  \bibinfo{author}{\bibfnamefont{A.}~\bibnamefont{Widera}},
  \bibinfo{author}{\bibfnamefont{T.}~\bibnamefont{M\"uller}}, \bibnamefont{and}
  \bibinfo{author}{\bibfnamefont{I.}~\bibnamefont{Bloch}},
  \bibinfo{journal}{Nature} \textbf{\bibinfo{volume}{448}},
  \bibinfo{pages}{1029} (\bibinfo{year}{2007}).

\bibitem[{\citenamefont{Trotzky et~al.}(2008)\citenamefont{Trotzky, Cheinet,
  F\"olling, Feld, Schnorrberger, Rey, Polkovnikov, Demler, Lukin, and
  Bloch}}]{trotzky_08}
\bibinfo{author}{\bibfnamefont{S.}~\bibnamefont{Trotzky}},
  \bibinfo{author}{\bibfnamefont{P.}~\bibnamefont{Cheinet}},
  \bibinfo{author}{\bibfnamefont{S.}~\bibnamefont{F\"olling}},
  \bibinfo{author}{\bibfnamefont{M.}~\bibnamefont{Feld}},
  \bibinfo{author}{\bibfnamefont{U.}~\bibnamefont{Schnorrberger}},
  \bibinfo{author}{\bibfnamefont{A.~M.} \bibnamefont{Rey}},
  \bibinfo{author}{\bibfnamefont{A.}~\bibnamefont{Polkovnikov}},
  \bibinfo{author}{\bibfnamefont{E.~A.} \bibnamefont{Demler}},
  \bibinfo{author}{\bibfnamefont{M.~D.} \bibnamefont{Lukin}}, \bibnamefont{and}
  \bibinfo{author}{\bibfnamefont{I.}~\bibnamefont{Bloch}},
  \bibinfo{journal}{Science} \textbf{\bibinfo{volume}{319}},
  \bibinfo{pages}{295} (\bibinfo{year}{2008}).

\bibitem[{\citenamefont{Giamarchi}(2004)}]{giamarchi_book}
\bibinfo{author}{\bibfnamefont{T.}~\bibnamefont{Giamarchi}},
  \emph{\bibinfo{title}{Quantum Physics in One Dimension}}
  (\bibinfo{publisher}{Clarendon Press}, \bibinfo{address}{Oxford},
  \bibinfo{year}{2004}).

\bibitem[{\citenamefont{Widera et~al.}(2008)\citenamefont{Widera, Trotzky,
  Cheinet, F\"olling, Gerbier, Bloch, Gritsev, Lukin, and Demler}}]{widera_07}
\bibinfo{author}{\bibfnamefont{A.}~\bibnamefont{Widera}},
  \bibinfo{author}{\bibfnamefont{S.}~\bibnamefont{Trotzky}},
  \bibinfo{author}{\bibfnamefont{P.}~\bibnamefont{Cheinet}},
  \bibinfo{author}{\bibfnamefont{S.}~\bibnamefont{F\"olling}},
  \bibinfo{author}{\bibfnamefont{F.}~\bibnamefont{Gerbier}},
  \bibinfo{author}{\bibfnamefont{I.}~\bibnamefont{Bloch}},
  \bibinfo{author}{\bibfnamefont{V.}~\bibnamefont{Gritsev}},
  \bibinfo{author}{\bibfnamefont{M.~D.} \bibnamefont{Lukin}}, \bibnamefont{and}
  \bibinfo{author}{\bibfnamefont{E.}~\bibnamefont{Demler}},
  \bibinfo{journal}{Phys. Rev. Lett.} \textbf{\bibinfo{volume}{100}},
  \bibinfo{pages}{140401} (\bibinfo{year}{2008}).

\bibitem[{blo()}]{bloch_private}
\bibinfo{note}{I. Bloch, private communication}.

\bibitem[{\citenamefont{Huber and Altman}(unpublished)}]{huber_09}
\bibinfo{author}{\bibfnamefont{S.~D.} \bibnamefont{Huber}} \bibnamefont{and}
  \bibinfo{author}{\bibfnamefont{E.}~\bibnamefont{Altman}},
  \bibinfo{journal}{arXiv:0907.0267}  (\bibinfo{year}{unpublished}).

\bibitem[{\citenamefont{Landau and Lifshitz}(1976)}]{landaulifshitzmec}
\bibinfo{author}{\bibfnamefont{L.~D.} \bibnamefont{Landau}} \bibnamefont{and}
  \bibinfo{author}{\bibfnamefont{E.~M.} \bibnamefont{Lifshitz}},
  \emph{\bibinfo{title}{Mechanics, Third Edition}}
  (\bibinfo{publisher}{Pergamon Press}, \bibinfo{year}{1976}).

\bibitem[{\citenamefont{Lieb and Liniger}(1963)}]{lieb1963}
\bibinfo{author}{\bibfnamefont{E.~H.} \bibnamefont{Lieb}} \bibnamefont{and}
  \bibinfo{author}{\bibfnamefont{W.}~\bibnamefont{Liniger}},
  \bibinfo{journal}{Phys. Rev.} \textbf{\bibinfo{volume}{130}},
  \bibinfo{pages}{1605} (\bibinfo{year}{1963}).

\bibitem[{\citenamefont{Rigol et~al.}(2007)\citenamefont{Rigol, Dunjko,
  Yurovsky, and Olshanii}}]{rigol2007}
\bibinfo{author}{\bibfnamefont{M.}~\bibnamefont{Rigol}},
  \bibinfo{author}{\bibfnamefont{V.}~\bibnamefont{Dunjko}},
  \bibinfo{author}{\bibfnamefont{V.}~\bibnamefont{Yurovsky}}, \bibnamefont{and}
  \bibinfo{author}{\bibfnamefont{M.}~\bibnamefont{Olshanii}},
  \bibinfo{journal}{Phys. Rev. Lett.} \textbf{\bibinfo{volume}{98}},
  \bibinfo{pages}{050405} (\bibinfo{year}{2007}).

\bibitem[{\citenamefont{Rigol}(2009)}]{rigol2009}
\bibinfo{author}{\bibfnamefont{M.}~\bibnamefont{Rigol}},
  \bibinfo{journal}{Phys. Rev. Lett.} \textbf{\bibinfo{volume}{103}},
  \bibinfo{pages}{100403} (\bibinfo{year}{2009}).

\bibitem[{\citenamefont{Tabor}(1989)}]{tabor1989}
\bibinfo{author}{\bibfnamefont{M.}~\bibnamefont{Tabor}},
  \emph{\bibinfo{title}{Chaos and Integrability in Nonlinear Dynamics}}
  (\bibinfo{publisher}{John Wiley {\&} Sons}, \bibinfo{year}{1989}).

\bibitem[{\citenamefont{Arnold}(1978)}]{arnold1978}
\bibinfo{author}{\bibfnamefont{V.~I.} \bibnamefont{Arnold}},
  \emph{\bibinfo{title}{Mathematical Methods of Classical Mechanics}}
  (\bibinfo{publisher}{Springer-Verlag}, \bibinfo{year}{1978}).

\bibitem[{\citenamefont{Arnold}(1963)}]{arnold1963}
\bibinfo{author}{\bibfnamefont{V.~I.} \bibnamefont{Arnold}},
  \bibinfo{journal}{Russ. Math. Surv.} \textbf{\bibinfo{volume}{18}},
  \bibinfo{pages}{85} (\bibinfo{year}{1963}).

\bibitem[{\citenamefont{Kolmogorov}(1957)}]{kolmogorov1957}
\bibinfo{author}{\bibfnamefont{A.~N.} \bibnamefont{Kolmogorov}},
  \bibinfo{journal}{Dokl. Akad. Nauk. SSSR} \textbf{\bibinfo{volume}{98}},
  \bibinfo{pages}{525} (\bibinfo{year}{1957}).

\bibitem[{\citenamefont{Moser}(1962)}]{moser1962}
\bibinfo{author}{\bibfnamefont{J.}~\bibnamefont{Moser}},
  \bibinfo{journal}{Nachr. Akad. Wiss. Goettingen Math. Phys.}
  \textbf{\bibinfo{volume}{K1}}, \bibinfo{pages}{1} (\bibinfo{year}{1962}).

\bibitem[{\citenamefont{Mazets et~al.}(2008)\citenamefont{Mazets, Schumm, and
  Schmiedmayer}}]{mazets}
\bibinfo{author}{\bibfnamefont{I.~E.} \bibnamefont{Mazets}},
  \bibinfo{author}{\bibfnamefont{T.}~\bibnamefont{Schumm}}, \bibnamefont{and}
  \bibinfo{author}{\bibfnamefont{J.}~\bibnamefont{Schmiedmayer}},
  \bibinfo{journal}{Phys. Rev. Lett.} \textbf{\bibinfo{volume}{100}},
  \bibinfo{pages}{210403} (\bibinfo{year}{2008}).

\bibitem[{\citenamefont{Kinoshita et~al.}(2006)\citenamefont{Kinoshita, Wenger,
  and Weiss}}]{kinoshita}
\bibinfo{author}{\bibfnamefont{T.}~\bibnamefont{Kinoshita}},
  \bibinfo{author}{\bibfnamefont{T.}~\bibnamefont{Wenger}}, \bibnamefont{and}
  \bibinfo{author}{\bibfnamefont{D.~S.} \bibnamefont{Weiss}},
  \bibinfo{journal}{Nature} \textbf{\bibinfo{volume}{440}},
  \bibinfo{pages}{900} (\bibinfo{year}{2006}).

\bibitem[{\citenamefont{Cazalilla}(2004)}]{cazalilla2004}
\bibinfo{author}{\bibfnamefont{M.}~\bibnamefont{Cazalilla}},
  \bibinfo{journal}{J. Phys. B} \textbf{\bibinfo{volume}{37}},
  \bibinfo{pages}{S1} (\bibinfo{year}{2004}).

\bibitem[{\citenamefont{Walls and Milburn}(1994)}]{walls-milburn}
\bibinfo{author}{\bibfnamefont{D.}~\bibnamefont{Walls}} \bibnamefont{and}
  \bibinfo{author}{\bibfnamefont{G.}~\bibnamefont{Milburn}},
  \emph{\bibinfo{title}{Quantum Optics}} (\bibinfo{publisher}{Springer-Verlag},
  \bibinfo{address}{Berlin}, \bibinfo{year}{1994}).

\bibitem[{\citenamefont{Gardiner and Zoller}(2004)}]{gardiner-zoller}
\bibinfo{author}{\bibfnamefont{C.}~\bibnamefont{Gardiner}} \bibnamefont{and}
  \bibinfo{author}{\bibfnamefont{P.}~\bibnamefont{Zoller}},
  \emph{\bibinfo{title}{Quantum Noise}} (\bibinfo{publisher}{Springer-Verlag},
  \bibinfo{address}{Berlin Heidelberg}, \bibinfo{year}{2004}),
  \bibinfo{edition}{3rd} ed.

\bibitem[{\citenamefont{Steel et~al.}(1998)\citenamefont{Steel, Olsen, Plimak,
  Drummond, Tan, Collett, Walls, and R.Graham}}]{steel}
\bibinfo{author}{\bibfnamefont{M.~J.} \bibnamefont{Steel}},
  \bibinfo{author}{\bibfnamefont{M.~K.} \bibnamefont{Olsen}},
  \bibinfo{author}{\bibfnamefont{L.~I.} \bibnamefont{Plimak}},
  \bibinfo{author}{\bibfnamefont{P.~D.} \bibnamefont{Drummond}},
  \bibinfo{author}{\bibfnamefont{S.~M.} \bibnamefont{Tan}},
  \bibinfo{author}{\bibfnamefont{M.~J.} \bibnamefont{Collett}},
  \bibinfo{author}{\bibfnamefont{D.~F.} \bibnamefont{Walls}}, \bibnamefont{and}
  \bibinfo{author}{\bibnamefont{R.Graham}}, \bibinfo{journal}{Phys. Rev. A}
  \textbf{\bibinfo{volume}{58}}, \bibinfo{pages}{4824} (\bibinfo{year}{1998}).

\bibitem[{\citenamefont{Polkovnikov}(2003{\natexlab{a}})}]{polkovnikov2003}
\bibinfo{author}{\bibfnamefont{A.}~\bibnamefont{Polkovnikov}},
  \bibinfo{journal}{Phys. Rev. A} \textbf{\bibinfo{volume}{68}},
  \bibinfo{pages}{053604} (\bibinfo{year}{2003}{\natexlab{a}}).

\bibitem[{\citenamefont{Blakie et~al.}(2008)\citenamefont{Blakie, Bradley,
  Davis, Ballagh, and Gardiner}}]{blakie_08}
\bibinfo{author}{\bibfnamefont{P.~B.} \bibnamefont{Blakie}},
  \bibinfo{author}{\bibfnamefont{A.~S.} \bibnamefont{Bradley}},
  \bibinfo{author}{\bibfnamefont{M.~J.} \bibnamefont{Davis}},
  \bibinfo{author}{\bibfnamefont{R.~J.} \bibnamefont{Ballagh}},
  \bibnamefont{and} \bibinfo{author}{\bibfnamefont{C.~W.}
  \bibnamefont{Gardiner}}, \bibinfo{journal}{Advances in Physics}
  \textbf{\bibinfo{volume}{57}}, \bibinfo{pages}{363} (\bibinfo{year}{2008}).

\bibitem[{\citenamefont{Polkovnikov}(2009)}]{polkovnikov2009}
\bibinfo{author}{\bibfnamefont{A.}~\bibnamefont{Polkovnikov}},
  \bibinfo{journal}{arXiv:0905.3384}  (\bibinfo{year}{2009}).

\bibitem[{\citenamefont{Polkovnikov et~al.}(2002)\citenamefont{Polkovnikov,
  Sachdev, and Girvin}}]{polkovnikov2002}
\bibinfo{author}{\bibfnamefont{A.}~\bibnamefont{Polkovnikov}},
  \bibinfo{author}{\bibfnamefont{S.}~\bibnamefont{Sachdev}}, \bibnamefont{and}
  \bibinfo{author}{\bibfnamefont{S.~M.} \bibnamefont{Girvin}},
  \bibinfo{journal}{Phys. Rev. A} \textbf{\bibinfo{volume}{66}},
  \bibinfo{pages}{053607} (\bibinfo{year}{2002}).

\bibitem[{\citenamefont{Polkovnikov}(2003{\natexlab{b}})}]{polkovnikov2003a}
\bibinfo{author}{\bibfnamefont{A.}~\bibnamefont{Polkovnikov}},
  \bibinfo{journal}{Phys. Rev. A} \textbf{\bibinfo{volume}{68}},
  \bibinfo{pages}{033609} (\bibinfo{year}{2003}{\natexlab{b}}).

\bibitem[{\citenamefont{Polkovnikov and Gritsev}(2008)}]{adiabnp}
\bibinfo{author}{\bibfnamefont{A.}~\bibnamefont{Polkovnikov}} \bibnamefont{and}
  \bibinfo{author}{\bibfnamefont{V.}~\bibnamefont{Gritsev}},
  \bibinfo{journal}{Nature Physics} \textbf{\bibinfo{volume}{4}},
  \bibinfo{pages}{477} (\bibinfo{year}{2008}).

\end{thebibliography}

\end{document}